\documentclass[english,reqno,12pt]{article}

\usepackage{graphicx,amsmath,amsfonts,verbatim,amssymb,amsthm}

\theoremstyle{plain}
\newtheorem{theorem}{Theorem}

\theoremstyle{remark}
\newtheorem{remark}{Remark}

\begin{document}

\title{Mathematical Conception of ``Phenomenological'' Equilibrium Thermodynamics}
\author{\bf V.~P.~Maslov}
\date{}

\maketitle

\begin{abstract}
In the paper, the principal aspects of the mathematical theory of equilibrium
thermodynamics are distinguished. It is proved that the points of degeneration
of a Bose gas of fractal dimension in the momentum space coincide with critical
points or real gases, whereas the jumps of critical indices and the Maxwell
rule are related to the tunnel generalization of thermodynamics. Semiclassical
methods are considered for the tunnel generalization of thermodynamics and also
for the second and ultrasecond quantization (operators of creation and
annihilation of pairs).
To every pure gas there corresponds a new critical point of the limit negative
pressure below which the liquid passes to a dispersed state (a foam). Relations
for critical points of a homogeneous mixture of pure gases are given in
dependence on the concentration of gases.
\end{abstract}

\begin{small}
\begin{align*}
\hbox to 6cm{}
&\text{"Leontii Sergeevich wrote a play himself!"}
\\
&\text{"What for?" asked Natalya Ivanovna with apprehension,}
\\
&\text{"Are we running out of plays? Such good plays around.}
\\
&\text{And so many\dots"}
\\
&\text{"Leontii wrote a modern play!"}
\\
&\text{Here the old lady became quite alarmed.}
\\
&\text{"We do not start mutinies against the authorities!"}
\\
&\qquad\qquad\qquad\qquad\qquad \textit{M.~Bulgakov ``Theatrical Novel''}
\end{align*}
\end{small}

\section{Introduction}

In his famous book on thermodynamics~\cite{1}, Leontovich refers to
thermodynamics as a phenomenological science, i.e., based on a series of
``statements established experimentally''. Beginning with the creation of
satellites and experiments in the absence of the gravity field of Earth, a new
period in physical experimental investigations of thermodynamical phenomena
occurred. Although these experiments are carried out by robots, and they cannot
react like a human being to any situation and to change the parameters of the
experiment like an animate observer, this must nevertheless unavoidably lead to
a reconsideration of the original conception of the great physicists.

On the other hand, computer-aided experiments have been developed so greatly
that a new science arose, the so-called molecular dynamics.

The problem of the mathematical axiomatization of thermodynamics and probability
theory was posed already by Hilbert.

Already Poincar\'e gave the standard definition of probability as the ratio of
the number of cases favorable for the event to the total number of possible
events\footnote{``La d\'efinition, dira-t-on, est bien simple: la probabilit\'e
d'un \'ev\'enement est le rapport du nombre de cas favorables \`a cet
\'ev\'enement au nombre total des cas possibles.''} and gave a counterexample
to this definition of probability. This definition must be completed, writes
Poincar\'e, by the sentence ``under the assumption that these cases are
equiprobable'' (\cite{2}, Russian p.~116) and notes that we have completed a vicious
circle place by defining probability via probability.\footnote{``On est donc
r\'eduit \`a compl\'eter cette d\'efinition en disant: `\dots au nombre des cas
possibles, pourvu que ces cas soient \'egalement probables.' Nous voil\`a donc
r\'eduits \`a d\'efinir le probable par le probable.''}

After this, Poincar\'e writes: ``The conclusion which seems to follow from
this\footnote{Poincar\'e presents a series of contradictions in probability
theory, including Bertrand's paradox.} is that the calculus of probabilities is
a useless science, that the obscure instinct which we call common sense, and to
which we appeal for the legitimisation of our conventions, must be
distrusted.'' (\cite{2}, Russian p.~116).\footnote{``La conclusion qui semble
r\'esulter de tout cela, c'est que le calcul des probabilit\'es est une science
vaine, qu'il faut se d\'efier de cet instinct obscur que nous nommions bon sens
et auquel nous demandions de l\'egitimer nos conventions.''}

Thus, first of all, the problem is to define what cases can be most naturally
assumed to be equiprobable. ``We are to look for a mathematical thought,''
writes Poincar\'e, ``where is remains pure, i.e., in arithmetic'' (\cite{24},
Russian p.~14).\footnote{``Il nous faut chercher la pens\'ee math\'ematique
l\`a o\`u elle est rest\'ee pure, c'est-\`a-dire en arithm\'etique.''}

The relationship between the quantum statistics of Bose gas with number theory
(``arithmetics'') was traced by Vershik, one of the greatest contemporary
mathematicians, who followed Temperley, a famous physicist~\cite{3}. Meanwhile,
Poincar\'e notes that one still cannot reject probability theory, because
probability theory is related to thermodynamics, and thermodynamics to
experiment.

The author had to study the modern experiments in the most thorough way and
profited from consultations with the great Russian virtuosi in experimental
physics, including V.~V.~Brazhkin, K.~I.~Shmulovich, A.~A.~Vasserman,
V.~I.~Nedostup, V.~G.~Baidakov, A.~E.~Gekhman, and also V.~A.~Istomin, the
greatest expert in applied thermodynamics and the theory of gas mixtures. It
was necessary to study the works of Professor~D.~I.~Ivanov especially
thoroughly; Professor Ivanov discovered a discrepancy between the
Widom--Kadanoff--Wilson theoretical critical
exponents~(\cite{4},~\cite{5},~\cite{6}) with the experimental
ones~(\cite{7},~\cite{8}) and carried out classical measurements in
thermodynamics.

Birkhoff, a great mathematician, devoted the first chapter of his book
``Hydrodynamics'' to paradoxes in hydrodynamics. In the strict mathematical
understanding, a paradox is a counterexample to a system of axioms.

First of all, the system of axioms we have constructed includes a modified
analog of the ``equidistribution law''~\cite{9}.

Our main axiom is the second one, which claims the (almost) uniform
distribution of densities in a vessel. This axiom leads to number theory,
where, instead of presenting all the theorems, we address the intuition of the
physicists who are acquainted with the Bose--Einstein distribution (this
distribution is closely related to number theory). This is sufficient for a new
conception of the notion of ideal gas. Further we use the third axiom, claiming
the existence of the Zeno-line for pure gases, which enables us to pass from
ideal gases to unperfect ones. To explain the Maxwell rule and the jumps of
critical indices at a critical point, as compared with the classical ones, we
use the axiom of tunnel quantization of the classical equation of state, which
thus encloses the system of four axioms.

The theory of ideal liquid treated as an incompressible one is a certain new
model in equilibrium thermodynamics.

\section{Heuristic considerations.\\ The role of small viscosity}

Following Clausius, experts in molecular physics usually argue by proceeding
from the symmetry of the motion of a molecule averaged in all six directions. In the
scattering problem, we use the principle of symmetry in all directions, which
is standard in molecular physics, but apply it to define not the mean free path,
but other molecular physics quantities. Therefore, the fraction of all
particles that moves head-on is 1/12. There are three such directions; hence,
one quarter of all molecules collide\footnote{The arguments put forward Clausius
concerning symmetry which were applied by Clausius to evaluate the free path length
and repeated here by the author are quite approximate. However, these arguments do
not influence on the values of ratios of the form $T_B/T_{\mathrm{cr}}$.}.

For the interaction potential, we consider the Lennard--Jones potential
\begin{equation}\label{1}
u(r)=4\varepsilon\Big(\frac{a^{12}}{r^{12}}-\frac{a^6}{r^6}\Big),
\end{equation}
where $\varepsilon$ is the energy of the depth of the well and $a$ is the
effective radius.

In the absence of an external potential, the two-particle problem reduces to
the one-dimensional radial-symmetric one. As is well known~\cite{10}, two
quantities: the energy~$E$ and the momentum~$M$ are conserved in this problem.
In the scattering problem, it is convenient to consider, instead of the
momentum~$M$, another preserved constant, namely, $b=\sqrt{m}\rho$, where
$\rho$ is the impact parameter, in such a way that
\begin{equation}\label{2}
M=\sqrt Eb.
\end{equation}

By solving with respect to energy $E$ the well-known relation
\begin{equation}\label{3}
E= \frac{p^2}{m} + \frac{M^2}{2mr^2}+u(r),
\end{equation}
we obtain the attracting Hamiltonian $H$:
\begin{equation}\label{4}
H=\frac{p^2/(2m)+u(r)}{1-b^2/r^2}, \qquad a<r\le b.
\end{equation}
The repulsing Hamiltonian is separated from $H$ by a barrier. Repulsive
particles put obstacles in the way of particles of the Hamiltonian $H$, by
creating ``viscosity''.

As the temperature decreases, the barrier height increases up to the value
$E_{\mathrm{cr}}=0.286\varepsilon$, and then starts decreasing (see~Fig.~1).
According to rough energy estimates~\cite{11}, for lesser temperatures, an
additional barrier must arise when clusters form. This barrier can be given
for carbon dioxide gas by micelles and for neutral gases and methane by germs
of droplets, i.e., three-dimensional clusters that contain at least one
molecule surrounded by other molecules (a prototype of a droplet) \footnote{By
a ``barrier'' we mean an obstacle for the collision of particles, namely, an
``envelope'' of surrounding particles that defends the given particle from an
immediate blow. In mathematics, a domain is an open region containing at least
one point.}.

\begin{figure}[h] %1
\begin{center}
\includegraphics{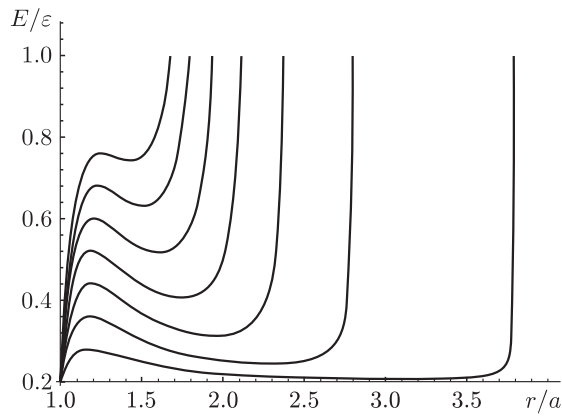}
\end{center}
\caption{The values of $E(b,r)$ for diverse values of the target parameter.}
\end{figure}

However, to study the penetration through the barrier of the incident particle,
we must plot~$E$ along the $y$ axis and turn the wells upside down. Then the
minimum becomes the barrier and the maximum becomes the depth of the well.

\begin{figure}[h] %2
\begin{center}
\includegraphics[width=7cm]{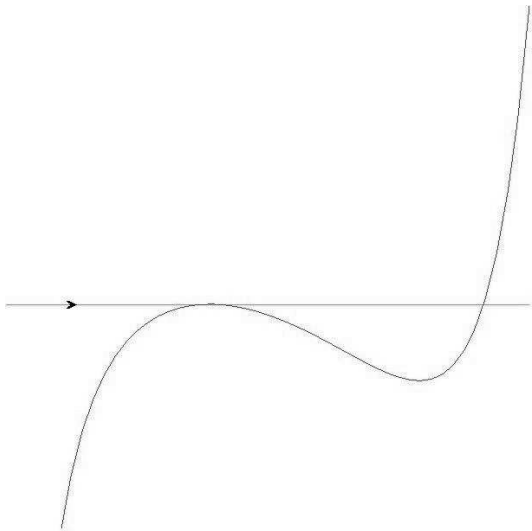}
\end{center}
\caption{The trap for a fictitious particle in the the center of mass coordinate system.
The radius vector $r$ of the $\mu$-point is marked on the abscissa axis. The
particle falls from the left from the point $r=\rho$, where $\rho$ stands for
the target parameter. In the original problem~\eqref{1} (before the change of
variables $M=\sqrt Eb$) the relation $r=\rho$ is attained as $r\to\infty$.}
\end{figure}

A dimer can be formed in a classical domain if the scattering pair has an
energy equal to the barrier height, slipping into the dip in ``infinite" time
and getting stuck in it as a result of viscosity (and hence of some small
energy loss), because this pair of particles, having lost energy, hits the
barrier on the return path. If the pair of particles has passed above this
point, then the viscosity may be insufficient for the pair to become stuck:
such a pair returns above the barrier after reflection. Therefore, only the
existence of a point~$E=E_{\max}$ plus an infinitesimal quantity, where
$E_{\max}$ is the upper barrier point, is a necessary condition for the pair to
be stuck inside the dip; $E_{\max}^{\mathrm{cr}}$ is the height of the maximum
barrier.

We can compare the values $T_{\mathrm{cr}}$ with the values
$E_{\max}^{\mathrm{cr}}$ in the table below.

\bigskip

$$
\begin{matrix}
&\text{Substance}\quad&\varepsilon, K \quad&{T_{cr}/4} \quad&
{E_{cr}\cdot\varepsilon/k}\\
&Ne\quad&36.3 \quad& 11 \quad& {10.5}\\
&Ar\quad&119.3 \quad& 37 \quad& {35}\\
&{Kr}\quad&171 \quad& 52 \quad& {50}\\
&{N_2} \quad& 95.9 \quad& 31 \quad& {28}\\
&{CH_4} \quad& 148.2 \quad& 47 \quad& {43}
\end{matrix}
$$

Above the value $E_{ \text{B}}=0.8\varepsilon$, the trap disappears. At the
value $0.286\varepsilon$, the depth of the trap is maximum and corresponds to
$T_{\mathrm{cr}}=\frac{1.16\varepsilon}k$. For neon and krypton, as can be seen
from the table, the concurrence is sufficiently good. Because
$T_B=3.2\varepsilon/k$, it follows that $T_B/T_{\mathrm{cr}}=2.7$, which
corresponds to the known relation of ``the law of corresponding
states''~\cite{12}.

The temperature corresponding to $4E_{ \text{B}}/k$, is the temperature above
which dimers do not appear. Exactly this is what we call the Boyle temperature
(in contrast to~\cite{9}).

In fact, an application of the Clausius approach to the pairwise interaction
gives a pairwise interaction with respect to the Lennard--Jones potential for
two Gibbs ensembles of noninteracting molecules. This leads to the presence of
a small friction for a single pair.

The difference $E_{\max} -E_{\min}$ is equal to the energy needed for a
particle lying at the bottom of the potential well to overcome the barrier. The
value $E_{\max}$ corresponds to the temperature given by $E_{\max}=RT$, where
$R$ stands for the universal gas constant. According to graph~2, $E_{\min}$
corresponds to the energy $PV$. Therefore, $E_{\min} /E_{\max} \leq 1$ is the
compressibility factor, $Z=PV/RT$. The temperature at the point
$E_{\min}=E_{\max}$ is equal to the Boyle temperature.

The dressed or ``thermal'' potential $\varphi(r)$ is attractive~\cite{13}. In
addition, because the volume~$V$ is a large parameter, it follows that, if the
quantity $\varphi(r)=({N\varepsilon a}/{\root3\of{V}})U\big(r/{\root3\of{V}}
\big),$ where $U\big(r/{\root3\of{V}}\big)$ is a smooth function and $N$ stands
for the number of particles, is expanded in terms of $1/\root3\of{V}$, then
\begin{equation}\label{5}
U\Big(\frac r{\root3\of{V}}\Big)=C_1+
\frac{C_2r}{\root3\of{V}}+\frac{C_3r^2}{(\root3\of{V})^2}
+O\bigg(\frac1{(\root3\of{V})^3}\bigg).
\end{equation}
Expanding
\begin{equation}\label{6}
C_1+r^2=\frac{(r-r_0)^2}2+\frac{(r+r_0)^2}2,
\end{equation}
where $C_1=r_0^2$, we can, just as in~\cite{13}, separate the variables in the
two-particle problem and obtain the scattering problem for pairs of particles
and the problem of their cooperative motion for $r_1+r_2$. The term
$C_2r/\root3\of{V}$ does not depend on this problem and the correction
$({a\varepsilon}/{\root3\of{V}})NO\big(1/V\big) $ is small.

Then, in the scattering problem, an attractive quadratic potential (inverted
parabola multiplied by the density or, to be more precise, by the
concentration, which we denote by the symbol~$\rho$ as well, because the target
parameter does not occur below) is added to
the Lennard--Jones interaction potential.

For this problem, we can find just as in~\eqref{3}--\eqref{4}, for all
$\rho=N/V$, a point corresponding to the temperature at which the well
capturing the dimers vanishes, and thus determine the so-called Zeno-line. It
is actually a straight line (up to 2\%), on which $Z=E_{\min}/E_{\max}=1$
(i.e., an ideal curve).

Let us clarify this fact in more detail.

We can treat the repulsing potential as a potential creating a small viscosity.

Let us find the total energy of the attracting Hamiltonian,
$$
E=\bigg(\frac{mv^2}{2(1-b^2/r^2)} \bigg)+\frac{\Phi(r)}{1-b^2/r^2}\,, \qquad
\Phi(r)=u(r)-\rho r^2.
$$
The first term is negative for $r\le b$ and the other term is positive for
$b>r>a$ (i.e., the more is the speed, the less is energy). The mean speed is
temperature.

Let us make the change of variables
$$
\frac ra=r', \qquad \frac ba=\widetilde b,
$$
and get rid of~$a$. In what follows, we omit both the tilde and the prime.

For a given $b$, the minimum $r_1$ and the maximum $r_2$
(see the graph no.~1 in~\cite{14}) are defined by the relation
\begin{equation}\label{7}
\frac{dE}{dr}=0.
\end{equation}

This gives $E_{\max}$ and $E_{\min}$. These values coincide at some point
$b=b_0$, and hence
\begin{equation}\label{8}
\frac{d^2E}{dr^2}=0
\end{equation}
at the point $r_0$, i.e., $E_{\max}=E_{\min}$, and this is the very Zeno-line.

Let us construct the curve $Z_{\min}=E_{\min}/E_{\max}$ minimal with respect to
the target parameter as a function of $\rho$. Let us find the point
$Z=E_{\min}/E_{\max}$ for $E_{\max}=E_{\max}^{\mathrm{cr}}$ and find the
corresponding point on the curve $Z_{\min}(\rho)$. This point is equal to
$Z_{\text{cr}}=0.29$, i.e., to the critical value of the compressibility
factor $Z$ for argon.

\begin{figure}[h] %3
\begin{center}
\includegraphics[width=7cm]{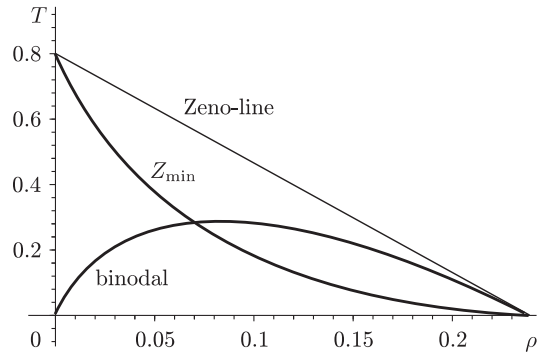}
\end{center}
\caption{The binodal, the Zeno-line, and the curve $Z_{\min}$. This heuristic binodal
does not coincide with the experimental one, whereas the Zeno-line and
$Z_{\mathrm{cr}}$ are close to the corresponding experimental curves.}
\end{figure}

In order to obtain a binodal according to some ``heuristic principle,'' we must
subtract the curve $Z_{\min}(\rho)$~\cite{15} from the Zeno-line. This gives
the graph shown in Fig~3.

\section{Quantum case}

In an example describing the creation of a dimer, it is shown that, for $T=T_{\mathrm{c}}$, one
degree of freedom becomes `` frozen'', and we obtain two degrees of freedom rather than three.
For a dimer with $T>T_{\mathrm{c}}$, if the oscillational degrees of freedom
are taken into account, then the number of degrees of freedom becomes equal to~6.
Two degrees of freedom are obtained under the assumption that
the oscillational degrees of freedom of the dimer are also ``frozen'' at $T=T_{\mathrm{c}}$.
If we suppose this heuristic supposition for the quantum case, then, for $T<T_d$,
both dimers with two degrees of freedom and dimers with six degrees of freedom are created.
This corresponds to the two-liquid Thiess--Landau model~[67].
In this case, the dimers with two degrees of freedom give the $\lambda $-point
and the dimers with six degrees of freedom give superfluidity.
Indeed, in the two-dimensional case we have
$$
c_p\cong\frac{2T}{T_d}\int_0^\infty\frac{\xi\,d\xi}{e^\xi-1}+\frac
T{T_d}\int_0^\infty\frac{e^\xi\xi\,d\xi}{(e^{\xi-\mu/T}-1)^2} +O\bigg(\frac{T-T_d}{T_d}\bigg),
$$
and we obtain a logarithmic divergence at the point $\xi=0$ for $\mu\to0$.

The case in which $N$ is not so large as it is in statistical physics, i.e.,
the so-called mesoscopic state (see~[68]), can also be of interest
for us. In this case, let us use Fock's idea for the Hartree equations, which
lead to the Hartree--Fock equations, which agree well with experiments
for a not too large number of electrons in atoms.

Namely, we consider the single-particle equation of the mean field (a
self-consistent field) and apply (to the resulting ``dressed'' potential) the
procedure of transition to the $N$-partial Schr\"odinger equation with a
dressed potential, just as we proceeded above for the operator $\frac{h^2}{2m}
\Delta$. Here we can consider two ways of investigation. The first way is the
way used by Fock and which leads in the semiclassical limit to the
Thomas--Fermi equations for the dressed potential. Another way is to consider
the Hartree temperature equations (see~[69]) and to obtain the
Thomas--Fermi temperature equations in the classical limit.

Since the quantity $T_d$ is small, it is easier to use the first way and to
find the ``dressed'' potential.

Let $V(q-q')$ be a pairwise interaction potential such that
$\int|V(r)|\,dr<\infty$. The dressed potential $W(q)$ is given by the formula
$$
W(q) =U(q)+ \int V(q-q')|\psi(q')|^2\,dq',
$$
where $U(q)$ stands for the external potential and $\psi(q')$ for an
eigenfunction of the Schr\"odinger equation which depends on the ``dressed''
potential and is thus an equation with a ``unitary'' nonlinearity. The
expansion of the equation in powers of $h$ can be found by the method of
complex germ up to $O(h^k)$, where $k$ is an an arbitrarily large
number\footnote{For $U(q)\equiv0$, one obtains Bogolyubov's famous
equation~[70]. The creation of dimers leads to the ultrasecond
quantization, i.e., to the operators of creation and annihilation of pairs.
This makes it possible to satisfy the boundary conditions in a
capillary~(46--48, 71--73).} (see~[69], where system~(63)
defines a complex germ; see also~[40], [75--80]).

Further, we consider two asymptotic formulas
for the number of eigenvalues of the one-particle Schr\"odinger operator.

The $2D$-dimensional phase space is partitioned into a lattice,
and the number $G_j$ is determined by the formula
\begin{equation}\label{l1}
G_i=\frac{\Delta p_j \Delta q_j}{(2\pi h)^D}.
\end{equation}
This relation was obtained by Weyl.

In the mathematical literature, the same notion for the number of eigenvalues
of the Laplace operator with the Dirichlet conditions
on the boundary of a $D$-dimensional volume,
which are less or equal to a given $\lambda$,
is called the Courant formula:
\begin{equation}\label{cour-1}
\rho(\lambda)= \frac{Vm^{D/2}\lambda^{D/2}}{\Gamma(D/2+1)(2\pi)^{D/2} h^D} (1+o(1))
\qquad \text{as} \quad \lambda\to\infty,
\end{equation}
where $D$ stands for the dimension of the space, because the spectral density
has the asymptotic behavior.

 The asymptotics
\begin{equation}\label{cour-2}
\lambda_j\sim \frac{2h^2}{m}\bigg(\frac{\pi^{D/2}\Gamma(D/2+1)}{V}\bigg)^{2/D} j^{2/D}
\qquad \text{as} \quad j\to\infty,
\end{equation}
is a natural generalization of formula~\eqref{cour-1}.

To reconcile the notion of Bose statistics which is given
in~[9] with symmetric solutions of the $N$-particle
Schr\"odinger equation, i.e., of the direct sum of $N$ noninteracting
Hamiltonians corresponding to the Schr\"odinger equation, and the symmetric
solutions of their spectrum, it is more appropriate to assign to the cells the
multiplicities of the spectrum of the Schr\"odinger equation in the way
described in~[81].

Consider the nonrelativistic case in which the Hamiltonian $H$ is equal to
${p^2}/({2m})$, where $p$ stands for the momentum.

The comparison of $G_i$ with the multiplicities of the spectrum of the
Schr\"odinger equation gives a correspondence between the eigenfunctions of the
$N$-partial Schr\"odinger equation that are symmetric with respect to the
permutations of particles and the combinatorial calculations of the Bose
statistics that are presented in~[9].

A single-particle $\psi$-function satisfies the free Schrodinger equation with
the Dirichlet conditions on the vessel walls.

Using precisely this very correspondence, we establish a relationship between
the Bose--Einstein combinatorics~[9],
the definition of the $N$-particle Schr\"odinger equation,
and the multiplicity of the spectrum of the single-particle Schr\"odinger equation.

The spectrum of the single-particle Schr\"odinger equation, provided that the
interaction potential is not taken into account, coincides, up to a factor,
with the spectrum of the Laplace operator. Consider its spectrum for the closed
interval, for the square, and for the $D$-dimensional cube with zero boundary
conditions. This spectrum obviously consists of the sum of one-dimensional
spectra.

On the line we mark the points $i=0,1,2,\dots$ and on the coordinate axes $x,y$
of the plane we mark the points with $x=i=0,1,2,\dots$ and $y=j=0,1,2,\dots$.
To this set of points $(i,j)$ we assign the points on the line that are
positive integers, $l=1,2,\dots$.

To every point we assign a pair of points, $i$ and $j$, by the rule $i+j=l$.
The number of these points is $n_l=l+1$. This is the two-dimensional case.

Consider the 3-dimensional case. On the axis $z$ we set $k=0,1,2,\dots$, i.e.,
let $$i+j+k=l$$ In this case, the number of points $n_l$ is equal to
$$
n_l=\frac{(l +1)(l +2)}{2}.
$$

It can readily be seen, for the $D$-dimensional case, that the sequence of
multiplicities for the number of variants $i=\sum_{k=1}^Dm_k$,
where $m_k$ are arbitrary positive integers, is of the form
\begin{equation}\label{3.7}
q_i(D) = \frac{(i+D-2)!}{(i-1)!(D-1)!}, \qquad \text{for} \ D=2, \quad q_i(2)=i,
\end{equation}
\begin{equation}\label{3.8}
\sum_{i=1}^\infty N_i=N, \qquad  \varepsilon \sum_{i=1}^\infty  q_i(D) N_i=E.
\end{equation}

The following problem in number theory corresponds to the three-dimensional
case $D=3$ (cf.~[9]):
\begin{equation}\label{dist_4}
\sum_{i=1}^\infty  N_i =N, \qquad
\varepsilon\sum_{i=1}^\infty \frac{(i+2)!}{i!6} N_i = M, \qquad  M=\frac{E}{\varepsilon}.
\end{equation}

In the manual~[9], following Einstein, a passage to the limit is
carried out as $N\to\infty$, which enables one to pass from sums to integrals.
Then, in the section ``Degenerate Bose gas,'' a point is distinguished which
corresponds to the energy equal to zero. This very point is the point of Bose
condensate on which excessive particles whose number exceeds some value
$N_d\gg1$ are accumulated at temperatures below the so-called degeneracy
temperature $T_d$. The theoretical discovery of this point anticipated a number
of experiments that confirmed this fact not only for liquid helium but also for
a series of metals and even for hydrogen.

From a mathematical point of view, distinguishing a point in the integral is an
incorrect operation if this point does not form a $\delta$ function.

The author obtained a more precise expression for the Bose--Einstein
but for finite values $N_d\gg1$ (see, e.g.,~[82]).

Thus, we consider the case in which $N\gg1$, but $n$ is not equal to infinity,
on the physical level of rigor~[82].
For the final parastatistics we have
\begin{equation}\label{kvas}
n_j= \frac{1}{\exp\{\frac{\varepsilon_j-\mu}{T}\} -1}
- \frac{k+1}{\exp\{(k+1)\frac{\varepsilon_j-\mu}{T}\} -1}, \qquad
n_j=\frac{N_j}{G_j}.
\end{equation}
In our case, we have $k=N_d$, and the point of condensate is $\varepsilon_0=0$.

By~\eqref{l1}, it is clear that $G_j$ is associated with the $D$-dimensional
Lebesgue measure and, in the limit with respect to the coordinates
$\Delta q_j$, gives the volume~$V$ in the space of dimension~$3$
and the area $Q$ in the space of dimension~$2$.
The passage with respect to the momenta $\Delta p_j$ is also valid
as $N\to\infty$ and $\mu>\delta>0$, where $\delta$ is arbitrarily small.

Expanding~(\ref{kvas}) at the point $\varepsilon_0=0$ in the small parameter
$x=(\mu N_d)/T_d$, where $N_d$ stands for the number of particles
corresponding to the degeneration and $T_d$ for the degeneracy temperature, and
writing $\xi=-\mu/T_d$, we obtain
\begin{gather}\label{dist1}
n_0= \bigg\{  \frac{1}{\exp\{\frac{-\mu}{T}\} -1}
- \frac{N_d+1}{\exp\{(N_d+1)\frac{-\mu}{T}\} -1} \bigg\}=
\frac{e^{\xi N_d}-1 -(N_d+1)(e^{\xi}-1)}{(e^{\xi}-1)(e^{2N_d}-1)}
\notag\\
=\frac{N_d}{2}\frac{1+\frac{x}{6}+\frac{x^2}{4!}+\frac{x^3}{5!}+ \dots}
{1+\frac{x}{2}+\frac{x^2}{6}+\frac{x^3}{4!}+ \dots}=
\frac{N_d}{2}\bigg(1-\frac{x}{3}-\frac{11}{24} x^2  -0.191 x^3-\dots\bigg).
\end{gather}

For example, if $x\to0$, then $n_0=N_d/2$, and hence the number $n_0$
in the condensate at $T=T_d$ does not exceed $N_d/2$.
If $x=1.57$, then $n_0\approx N_d/10$. Certainly, this affects the degeneracy temperature,
because this temperature can be expressed only in terms of the number of
particles above the condensate, $\tilde{N}_d$, rather than in terms of the
total number of particles $N_d$ (which is equal to the sum of $\tilde{N}_d$ and
of the number of particles in the condensate).

Write $M=E_d/\varepsilon_1$, where $\varepsilon_1$ stands for the coefficient
in formula~\eqref{cour-2} for $j=1$. Let us find $E_d$,
\begin{equation}\label{cour-3}
E_d= \int_0^\infty \frac{\frac{|p|^2}{2m}\,
d\varepsilon}{e^{\frac{|p|^2}{2m}/T_d}-1},
\end{equation}
where
\begin{equation}\label{cour-4}
d\varepsilon=\frac{|p|^2}{2m} \frac{dp_1 \dots dp_D \, dV_D}{(2\pi h)^D}.
\end{equation}
Whence we obtain the coefficient $\alpha$ in the formula,
\begin{equation}\label{cour-5}
E_d =\alpha T_d^{2+\gamma}\zeta(1+D/2)\Gamma(1+D/2).
\end{equation}

To begin the summation in~\eqref{3.8} at the zero index (beginning with the
zero energy), it is necessary to rewrite the sums~\eqref{3.8} in the form
\begin{equation}\label{3.9}
\sum_{i=0}^\infty N_i=N, \qquad
\varepsilon \sum_{i=0}^\infty (q_i(D)-1) N_i=E-\varepsilon N.
\end{equation}

The relationship between the degeneracy temperature and the number
$\tilde{N}_d$ of particles above the condensate for $\mu>\delta>0$ (where
$\delta$ is arbitrarily small) can be found for $D>2 $ in the standard way.

Thus, we have established a relationship between $G_i$ in formula~\eqref{l1}
(which is combinatorially statistical) and the multiplicity of the spectrum for
the single-particle Schr\"odinger equation, i.e., between the
statistical~[9] and quantum-mechanical definitions of Bose particles.

Consider the two-dimensional case in more detail.
There is an Erd\H os' theorem for a system of two Diophantine equations,
\begin{equation}\label{df-eq}
\sum_{i=1}^\infty N_i=N, \qquad  \sum_{i=1}^\infty i N_i=M.
\end{equation}
The maximum number of solutions of this system is achieved
if the following relation is satisfied:
\begin{equation}\label{108}
N_d=c^{-1}M_d^{1/2} \log\, M_d+ aM_d^{1/2}+o(M_d^{1/2}), \qquad
c=\pi\sqrt{2/3},
\end{equation}
and if the coefficient $a$ is defined by the formula $c/2=e^{-ca/2}$.

The decomposition of $M_d$ into one summand gives only one version. The
decomposition $M_d$ into $M_d$ summands also provides only one version (namely,
the sum of ones). Therefore, somewhere in the interval must be at least one
maximum of the variants. Erd\H os had evaluated it~\eqref{108}
(see~[83]).

If the number $N$ increases and $M$ is preserved in the problem~\eqref{df-eq},
then the number of solutions decreases. If the
sums~\eqref{df-eq} are counted from zero rather than from one, i.e., if we set
\begin{equation}\label{106}
\sum_{i=0}^\infty  i N_i = (M-N), \qquad
\sum_{i=0}^\infty  {N_i} = N,
\end{equation}
then the number of solutions does not decrease and remains constant.

I shall try to explain this effect. The Erd\H os--Lehner problem~[84]
is to decompose $M_d$ into $N\leq N_d$ summands.

Let us expand the number $5$ into two summands. We obtain $3+2=4+1$.
The total number is $2$ versions (this problem is known as ``partitio numerorum'').
If we include~$0$ to the possible summands, we obtain three versions:
$5+0=3+2=4+1$. Thus, the inclusion of zero makes it possible to say
that we expand a number into $k\leq n$ summands.
Indeed, the expansion of the number $5$ into three summands includes
all the previous versions, namely, $5+0+0$, $3+2+0$, and $4+1+0$,
and adds new versions, which do not include zero.

In this case, the maximum does not change drastically~[84];
however, the number of versions is not changed, namely, the zeros, i.e.,
the Bose condensate, make it possible that the maximum remains constant,
and the entropy never decreases; after reaching the maximum, it becomes constant.

Let us turn to a physical definition.

Note first that, without changing the accuracy of the quantity whose logarithm
is evaluated, we can replace $\log M_d$ by $(1/2)\log ({\tilde{N}_d}/{Q})$.
Then
\begin{equation}\label{cour-6}
\sqrt{M_d} = \frac{2 \tilde{N}_d/Q}{c^{-1} \log (\tilde{N}_d/Q) +a}+
o\bigg(\frac{\tilde{N}_d}{Q}\bigg).
\end{equation}
In our case, $\tilde{N}_d/Q$ corresponds to the number of particles above the
condensate.

According to formula~\eqref{cour-5}, in the two-dimensional case we must set
$\gamma=0$ and find the coefficient $\alpha$. Then formula~\eqref{cour-6} gives
us a relationship between $\tilde{N}_d$ and $T_d$ due to the fact that the
number of particles in the condensate is $o(\tilde{N}_d)$.

In fact, we have proved that there is a gap between $\mu>\delta>0$
and $\mu=0$.

\section{Main axiom and theorem for the classical gas}

Consider the Maxwell distribution
\begin{equation}\label{9}
\omega(p)=\frac1{(2\pi mT)^{3/2}}\exp\Big\{-\frac{p^2}{2mT}\Big\}.
\end{equation}
As is well known\footnote{We have kept the integral especially without
evaluating it.}, it is associated with the potential
\begin{equation}\label{10}
\Omega=\frac{V\pi^{3/2}T^{5/2}m^{5/2}}{\Gamma(2+1/2)}\int_0^\infty
t^{3/2}e^{-(t-\frac\mu T)}\,dt.
\end{equation}

Now consider an ultrarelativistic gas in which the kinetic energy is
proportional to~$|p|$. In this case,
\begin{equation}\label{11}
\omega(p)=\frac{c^3}{(\pi T)^3}\exp\Big\{-\frac{c|p|}T\Big\},
\end{equation}
where $c$ is the velocity of light. For this gas, the potential~$\Omega$ is of
the form
\begin{equation}\label{12}
\Omega={VT^3}\frac{\pi^2}{\Gamma(3)c^3}\int_0^\infty t^3e^{-(t-\frac\mu T)}\,dt.
\end{equation}
Since $N=-\frac{\partial\Omega}{\partial\mu}$ and
$P=-\frac{\partial\Omega}{\partial V}$, it follows that, in both the first and
the second case, we obtain the equation of an ideal gas
\begin{equation}\label{13}
PV=NT.
\end{equation}

In thermodynamics, there is an important notion, namely, $l$, the number of
the degrees of freedom of the molecule. The energy of the molecule is
$$
\text{ $E\sim l\frac{p^2}{2m}$\,.}
$$
The parameter $l$ (the number of degrees of freedom) is the coefficient at the
potential~$\Omega$. It does not influence the equation of ideal gas~\eqref{13}.
This coefficient is related to the so-called ``equidistribution law'' mentioned
above. Possibly, the law is true in some interval of temperatures. However, as
the temperature reduces. some degrees of freedom can be ``frozen.'' It turns
out that it is much more convenient to introduce the parameter $\sigma$
presented below.

Assuming that the value of $l$ depends on the momentum, we replace the above
relation by the following one:
\begin{equation}\label{14}
E_\gamma=\frac{|p|^{2+\sigma}}{2mp_0^{\sigma}},
\end{equation}
where $p_0$ stands for some typical momentum.

We can say that both the average momentum and the number of the degrees of
freedom depend on the temperature, and hence the number of the degrees of
freedom depends on the momentum. We assume that this dependence is the simplest
one, namely, it is a power dependence (the parameter~$\sigma$ of in the
exponent characterizes the molecule). Here the potential $\Omega$ takes the
form
\begin{gather}\label{15}
\Omega_\gamma= \frac{\pi^{1+\gamma}}{\Lambda^{2(1+\gamma)}} VT^{2+\gamma}
\frac{m^{2+\gamma}}{\Gamma(2+\gamma)} \int_0^\infty  t^{1+\gamma}
e^{-(t-\frac{\mu}{T})}\, dt,\\
\gamma=\frac{3}{2+\sigma}-1=\frac{1-\sigma}{2+\sigma},
\nonumber
\end{gather}
where $\Lambda$ stands for a constant corresponding to a given molecule and
depending on its mass (however, we try to avoid the mass by passing from the
density to the concentration). It can readily be seen that, in this case,
relation~\eqref{13} remains valid for an ideal gas.

Thus, we replace the parameter of the integer degrees of freedom by some
continuous parameter that characterizes the given molecule. In principle, this
parameter $\gamma$ is of the same physical origin as the number of the degrees
of freedom. But since it is continuous, it takes into account more details of
the spectrum of the molecule.

Moreover, below we shall see the following remarkable fact: every molecule is
characterized by its own value of $\gamma$ which does not depend on temperature
and depends only on the dimensionless value $Z_{\mathrm{cr}}$, which is its
critical compressibility factor.

One must not think that this relation (formula~\eqref{14}) holds for a single
molecule. It can be treated only as a result of averaging over the Gibbs
ensemble of noninteracting molecules that corresponds to some temperature (as
we shall see below, this ensemble corresponds to the critical temperature
$T_{\mathrm{cr}}$).

Roughly speaking, the number of degrees of freedom is related to the number of
atoms~\cite{9}, whereas the number of atoms is related to the dimension of the
space. For this reason, the reduction of the number of degrees of freedom, as
the temperature reduced, is equivalent in a Gibbs ensemble of noninteracting
molecules to the reduction of the mean value of the space dimension. On the
other hand, one can say that the reduction of the number of degrees of freedom
is analogous to the reduction of the dimension of the space of oscillators
(from three-dimensional to zero-dimensional oscillators), whereas the latter
reduction is equivalent (see Sec.~5) to the halving of the dimension of the
momentum space. However, for the value of the potential $\Omega$, increasing of
the exponent of $p$ in energy is equivalent to the reduction of the fractal
dimension when integrating with respect to $p$. Therefore, the passage
to~$E(p)=ap^{\alpha}$, which is regarded as an equivalent notion of fractal
dimension, is related to averaging of the Gibbs ensemble of noninteracting
molecules with respect to the number of active degrees of freedom. Therefore,
the formal change of variable in manuals~(\cite{60}, p.~212) and
handbooks~\cite{61} (see the section ``Integral representations'' in~\cite{61}
and also the book~\cite{62}) has a well-founded physical meaning of continuous
reduction of a fractional value of the number of degrees of freedom of a
molecule as the temperature reduces (in the mean) for a Gibbs ensemble of
noninteracting molecules.

A.~V.~Chaplik suggested another interpretation of this law. Since it turns out
that a law of the form $ap^{\alpha}$ enables us to describe the experiment, and
since we assume that this is an ideal gas of particles with this dependence,
$E(p)=ap^{\alpha}$, it follows that these are quasiparticles, i.e., collective
excitements (rather than individual molecules). The $p$-momentum of the
quasiparticle is preserved, because the interaction between the particles is no
longer valid. This trick is customary in physics.

However, in mathematics, the notion of Gibbs ensemble corresponds to the notion
of the Kolmogorov complexity, and therefore, for a mathematician, it is more
natural to relate the formula $E(p)=ap^{\alpha}$ to this vary complexity.
However, the main point is that this generalizes the notion of equidistribution
to lower temperatures.

Now we can rigorously formulate the axiom of thermodynamics corresponding to
the approximate conservation of the gas density (this corresponds to the
physicists' statement in equilibrium thermodynamics: ``the density is
homogeneous in a vessel''; physicists consider equilibrium thermodynamics as a
separate discipline, after which fluctuation theory is also considered
separately).

A physicist thinks that the density of particles inside the vessel is constant
up to fluctuations (of the order of the square root of the number of particles
in a small subvolume). This means that the physicist counts the number of
particles in every small subvolume containing approximately one million of
particles, and the density does not depend on the permutation of the numbers of
particles. However, this is arithmetics, in which the sum does not depend on
the permutation of summands (in contrast to the Boltzmann statistics in which a
permutation of particles gives a new state). Even if we assume that the
experimenter had indexed all particles during the previous measurement, this
experimenter cannot control what index corresponds to a given particle at the
next moment of measurement, and the indexing can be carried out again to
compute the density. This is a fact in arithmetics, which is the very base of a
special branch of science, namely, of analytic number theory.

\medskip

{\bf Main mathematical axiom of thermodynamics.} {\it Consider a vessel of
volume $V$ containing $N>10^{19}$ identical molecules corresponding to the
parameter $\gamma=\gamma_0$. Consider a small convex volume of size~$V_\delta$
containing $N_\delta$ particles, where $N_\delta$ is not less than $10^6$. Let
$\mathbb P$ be the probability of the deviation $$N_\delta\pm\sqrt{N_\delta}$$
for any volume of size~$V_\delta$ inside the vessel. The claim of the axiom is
that the probability~$\mathbb P$ is sufficiently small. Namely,}
\begin{equation}\label{16}
\mathbb P\biggl\{\biggl|\frac NV-\frac{N_\delta}{V_\delta}\biggr|
>\frac{\sqrt{N_\delta}}{V_\delta}\biggr\}
\le\frac{0.01}{\sqrt{\log N_\delta}}\,.
\end{equation}

This implies that the numbering of particles in the volume $V_\delta$ is
arbitrary and does not affect the density. We can rearrange the numbers, and
this also does not affect the density, because the rearrangement of summands
does not change the sum\footnote{The set of initial data for an $N$-particle
dynamical system can be said to be chaotic if a relation of the form~\eqref{16}
arises at a sufficiently large time interval. The weakening of
condition~\eqref{16} does not influence the validity of relation~\eqref{18}.}.

\begin{remark} \rm
Condition~\eqref{16} shows that the theory presented
below cannot be applied for insufficiently many particles. This condition also
shows what is the accuracy that can be expected from the relations obtained
below. The weakening of condition~\eqref{16} plays no fundamental role, and is
manifests itself only at the level of the expected accuracy of these relations.
\end{remark}

Thus, we have already proved that the approximate conservation of density
inside a gas volume implies that the particles are independent of the
numbering, and hence are Bose particles. Therefore, the potential
$\Omega_\gamma$~\eqref{15} must be replaced by the Bose gas
potential\footnote{The Bose gas is usually regarded as a quantum gas;
however, it is related to number theory (see Remark~2 below),
and number theory is related to the main axiom.
Therefore, the potential $\Omega_\gamma$~\eqref{17} can be
applied to classical gas. In number theory, $V\equiv1$. To make the
perception by the physicists more convenient, we keep the multiplier $V$
(cf.~\eqref{21}--\eqref{22} below).},
\begin{equation}\label{17}
\Omega_\gamma= \frac{\pi^{1+\gamma}}{\Lambda^{2(1+\gamma)}} VT^{2+\gamma}
\frac{1}{\Gamma(2+\gamma)} \int_0^\infty  \frac{t^{1+\gamma}}{
e^{(t-\frac{\mu}{T})}-1}\, dt = \frac{\pi^{1+\gamma}}{\Lambda^{2(1+\gamma)}}
VT^{2+\gamma}\operatorname{Li}_{\gamma+2}(e^{\mu/T}),
\end{equation}
where $\operatorname{Li}_\gamma(\cdot)$ stands for the polylogarithm,
$\operatorname{Li}_{\gamma+2}(1)=\zeta(\gamma+2)$, and $\zeta$ stands for the
Riemann zeta function.

Thus, in our case, the Bose gas also depends on another parameter, $\gamma$. We
refer to this parameter as the {\it fractional $($``fractal''$)$ dimension} in the
momentum space (cf.~\cite{9}, Sec.~Degenerate Bose gas, footnote). the
parameter could be called more precisely the {\it parameter characterizing
the spectrum of the molecule}.

\begin{remark} \rm
In the statistics of Boltzmann, Shannon, and Kolmogorov,
$4+1$ and $1+4$ are different variants. In arithmetic, they constitute one
variant. For $\gamma>0$, the Bose statistics is related to number theory. In
the physical literature, this was first noted by Temperley~\cite{3}; in the
mathematical literature, this was considered in detail in~\cite{16}--\cite{18}.
\end{remark}

\begin{remark} \rm
Note that, just as for the standard potential
$\Omega_\gamma$~\eqref{15}, three-dimensional Bose gas corresponds to
$\gamma=1/2$, whereas $\gamma=2$ corresponds to ultrarelativistic gas. The
latter is associated with the famous Planck distribution at $\mu=0$ for the
black-body radiation (cf.~\cite{9}, Sec.~60).
\end{remark}

\begin{theorem}
The point of degeneracy of the Bose distribution at
$\mu=0$ for molecules of characteristic~$\gamma$ coincides with the critical
points of these molecules up to normalization. Therefore, the dimensionless
quantity
$$
Z_{\mathrm{cr}}=\frac{V_{\mathrm{cr}}P_{\mathrm{cr}}}{R T_{\mathrm{cr}}},
$$
where $R$ stands for the gas constant, coincides with the dimensionless point
of degeneracy of an ideal Bose gas of fractal characteristic~$\gamma$. Hence,
\begin{equation}\label{18}
Z_{\mathrm{cr}}= \bigg(\frac VT\frac{\partial\Omega_\gamma}{\partial
V}\bigg/\frac{\partial\Omega_\gamma}{\partial\mu}\bigg)_{\mu=0}
=\frac{\zeta(\gamma+2)}{\zeta(\gamma+1)},
\end{equation}
which is a consequence of the number-theoretic interpretation of the Bose
distribution.
\end{theorem}

This is a property of the new ideal gas without interaction. Apfelbaum and
Vorob'ev~\cite{19} computed the critical isotherms in the $\{P,V\}$-plane for
different gases, where the value of~$\gamma$ is determined by the above
formula, and the critical isotherm of the Bose gas with the evaluated
characteristic~$\gamma$ was computed by the relevant formulas for an ideal Bose
gas \footnote{This gas is more general than that in~\cite{20} and corresponds
to the fractal dimension~$\gamma$ in the momentum space. In~\cite{20},
$\gamma=1/2$. There is no classical gas of this dimension.} \ (see Figs.~4--5).

\begin{figure}[h] %4
\begin{center}
\includegraphics{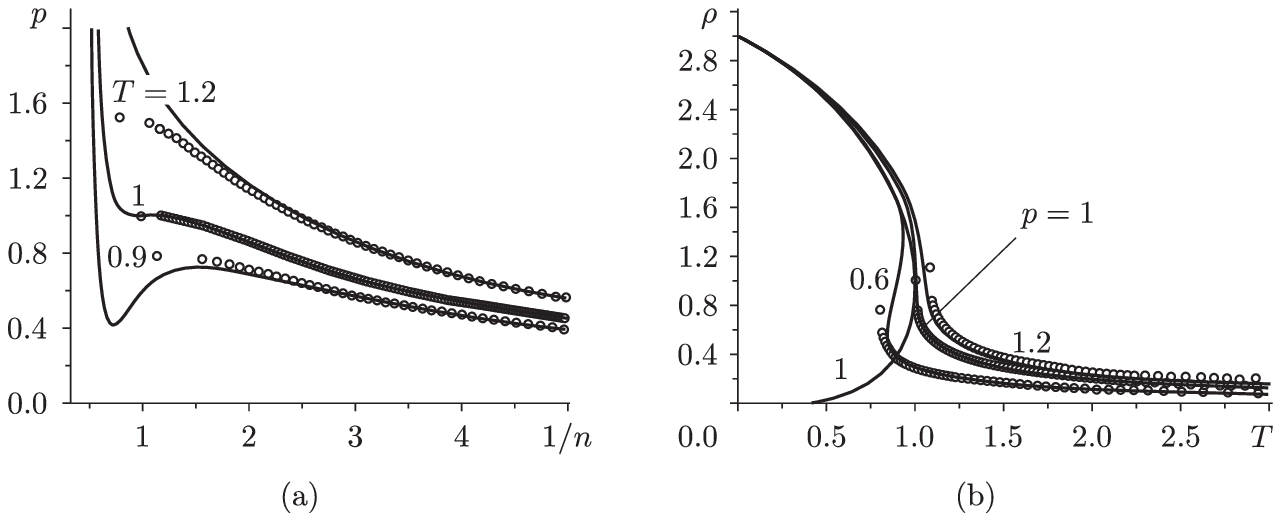}
\end{center}
\caption{(a) Isotherms of pressure for the van der Waals equation are
shown by continuous lines. The small circles show the corresponding lines
computed with $\gamma=0.312$ for $\varphi(V)=V$ (i.e., ideal Bose gas),
$Z_{\mathrm{cr}}=3/8$.
\newline
(b) Isobars of density for the van der Waals equation are shown by continuous
lines. Line 1 is the binodal curve. The small circles correspond to
Bose--Einstein isobars for $\gamma=0.312$.}
\end{figure}

\begin{figure}[h] %5
\begin{center}
\includegraphics{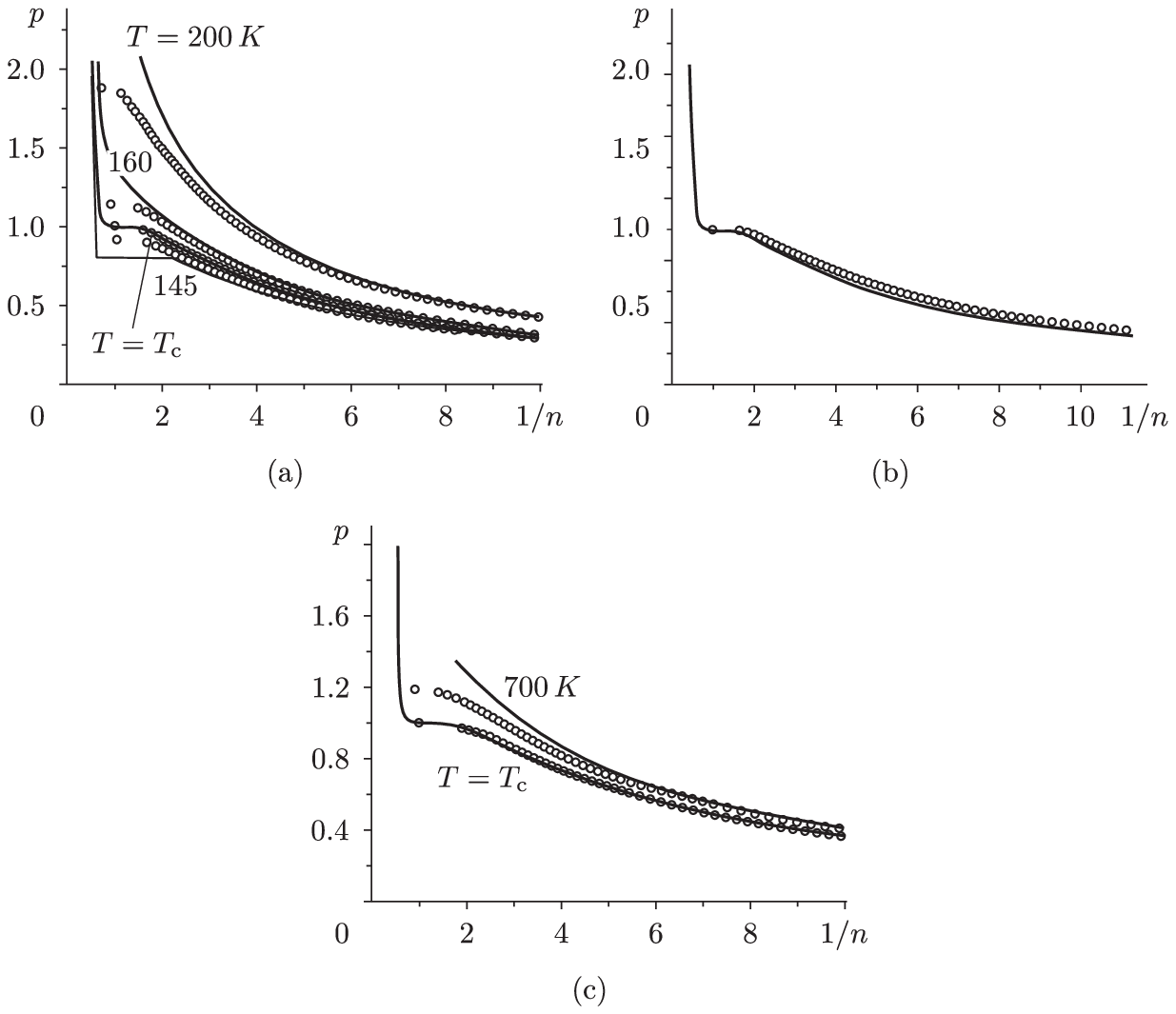}
\end{center}
\caption{(a) Isotherms for argon. The continuous lines correspond to the experimental
data, and the line formed by small circles is constructed according the
isotherm of ideal Bose gas.
\newline
(b) Isotherms for water. $Z_{\mathrm{cr}}=0.23$.
\newline
(c) Critical isotherms for copper. $Z_{\mathrm{cr}}=0.39$.}
\end{figure}

At first glance, it looks as if the notion of new ideal gas leads to an
alteration of the famous relation
\begin{equation}\label{19}
PV=NT
\end{equation}
(which, moreover, served as an analogy for the main economics law, the Irving
Fisher formula; which is used to calculate the ``turnover rate'' of
capital~\cite{21}). This would be surprising indeed. However, this is not the
case. The relation $PV=NT$ or, equivalently, $PV=RT$ (because the number of
particles in the vessel remains the same) defines an imperfect gas and, in
contemporary experimental thermodynamical diagrams, it is called the
{\it Zeno-line} or, sometimes, the ideal curve, the Bachinskii parabola, etc.

On the diagram $(\rho,T)$ for pure gases, this is the straight line $Z=1$. The
line is a most important characteristic feature for a gas which is imperfect.
Since, for imperfect gases, it has been calculated experimentally and is an
``almost straight'' line on the $(\rho,T)$ diagram, it follows that the
Zeno-line is determined by the two points $T_B$ and $\rho_B$ called the ``Boyle
temperature'' and the ``Boyle density.'' In contrast to $Z_{\mathrm{cr}}$,
these points are related to the interaction and scattering of a pair of gas
particles with the interaction potential peculiar to this gas, as was shown in
Sec.~1 and in other papers of the author (see, e.g.,~\cite{22}). Therefore, the
Zeno-line on which the relations
\begin{equation}\label{20}
PV=NT, \qquad \frac\rho{\rho_B}+\frac T{T_B}=1, \qquad \frac NV=\rho,
\end{equation}
are satisfied, where $\rho$ is the density (the concentration), is a
consequence of pairwise interaction, and thus is a relation for an imperfect
gas.

The correction related to the existence of the Zeno-line leads to a
differential equation~\cite{23} whose numerical solution yields an alteration
to the gas spinodal for every particular pure gas. For argon and $CO_2$, this
modification is shown in Fig.~6\footnote{By the heuristic considerations
related to the scattering problem (Sec.~1), the final point~\cite{24} of the
gas spinodal is equal to $Z=3/2\,Z_{\mathrm{cr}}$, and the spinodal can be
approximated by a line segment. (Ideally, in infinite time, a fictitious
particle (a pair) falls to the bottom due to friction, i.e., the orbit of this
particle is circular, and thus one degree of freedom disappears. This means
that the compressibility factor $Z=0.444$ at the point $E_{\max}^{\text{cr}}$
is reduced by the factor 2/3, i.e., $Z_{\text{cr}}=0.296$.). This makes it
possible to construct two points near $P=0$, $Z=Z_{\mathrm{cr}}$ by the
theories of a new ideal (Bose) gas and by the fact that the chemical potential
of the gas is equal to the chemical potential of an ideal liquid, and thus to
approximately reconstruct the Zeno-line.}.

The distribution of number theory, as opposed to the Bose--Einstein
distribution~\eqref{17}, does not contain the volume $V$. Let us consider the
distribution of number theory multiplied by unknown
function~$\varphi_{\gamma_0}(V)$ which does not vary for $\gamma\ge\gamma_0$
and $T\le T_{\mathrm{cr}}$. Then it follows from~\eqref{20} that
\begin{gather}\label{21}
P=\frac{\varphi'_{\gamma_0}(V)T^{\gamma_0+2}}{\Gamma(\gamma_0+2)}
\int^\infty_0\frac{\varepsilon^{\gamma_0+1}\,d\varepsilon}
{e^{-\kappa}e^\varepsilon-1}, \qquad
\varphi'_{\gamma_0}=\frac{d\varphi_{\gamma_0}}{dV},
\nonumber\\
\varphi_{\gamma_0}'(V)\operatorname{Li}_{\gamma_0+2}(y)=
\frac\rho{T_B^{\gamma_0+1}\big(1-\frac\rho{\rho_B}\big)^{\gamma_0+1}}, \quad
\rho=\frac RV, \quad y=e^\kappa, \quad \kappa=\frac\mu T.
\end{gather}

The differential equation for $\varphi_{\gamma_0}$ is
\begin{equation}\label{22}
\frac{V\varphi'_{\gamma_0}(V)}{\varphi_{\gamma_0}(V)}
\frac{\operatorname{Li}_{\gamma_0+2}(y)} {\operatorname{Li}_{\gamma_0+1}(y)}=1,
\qquad V=\frac R\rho.
\end{equation}
See Fig.~6.

\begin{figure}[h] %6
\begin{center}
\includegraphics{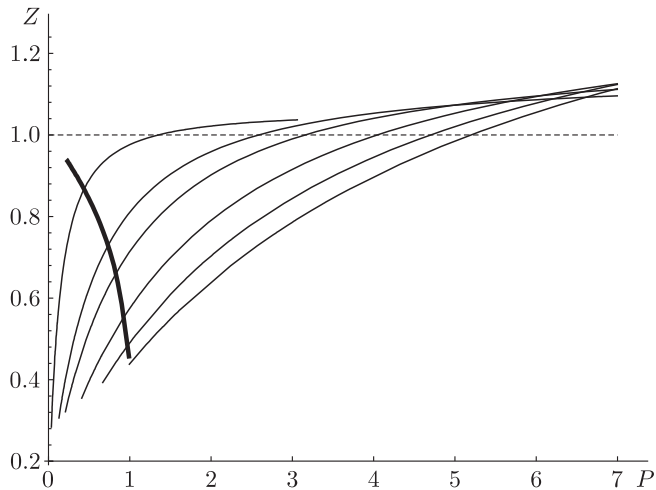}
\end{center}
\caption{The dotted line shows the Zeno-line $Z=1$. The bold line is the critical
isotherm of an imperfect gas (argon) calculated theoretically; the thin lines
correspond to the isochores of an imperfect gas for $T<T_{\mathrm{cr}}$. Their
initial points lie on the gas spinodal.}
\end{figure}

\section{Ideal liquid}

Let us now pass to the notion of ideal liquid. For an expert in mathematical
physics, an ideal liquid is an incompressible liquid. In our mathematical
conception of thermodynamics, we shall abide by this definition. In this case,
on the Zeno-line on the plane $\{P,Z\}$, the point $P(T,\rho)$ in~\eqref{20} is
uniquely defined for $Z=1$. The isotherm $T= \text{const}$ is a straight line.
The second point is a point of the spinodal.

As is well known, the passage from the gaseous state to the liquid one is
accompanied by the entropy drop. Naturally, the entropy, which determines the
measure of chaotic behavior, is less for the liquid state than for the gaseous
state. At the same time, the general property of ``choosing'' a subsystem with
the greatest chaoticity among all possible subsystems leads to the property of
constant entropy of the liquid, which was noted both theoretically and
experimentally, even if the temperature tends to the absolute
zero~(\cite{63},~\cite{64}) (the entropy tends to $\log 2$).

In our model of ideal liquid as an incompressible liquid, we suppose in
addition that the maximum of the entropy on a given isotherm (i.e., as
$\mu \to0$) does not vary when the temperature varies (see below Section~6).

The big thermal potential is of the form
\begin{equation}\label{23}
\Omega= -PV_{\gamma}=- \frac{\pi^{\gamma+1}V_{\gamma}T}{\Lambda^{2(1+\gamma)}}
\frac{1}{\Gamma(2+\gamma)}\int_0^\infty \frac{t^{1+\gamma}\,dt}{(e^t/z)-1}=
\frac{-\pi^{\gamma+1}L^{2(1+\gamma)}T^{2+\gamma}}{{\Lambda}^{2(1+\gamma)}}
\operatorname{Li}_{2+\gamma}(z).
\end{equation}

The entropy becomes
\begin{align}\label{24}
S=-\bigg(\frac{\partial\Omega}{\partial T}\bigg)_{V,\mu}&
=(2+\gamma) \frac{L^{2(1+\gamma)}T^{1+\gamma}}{{\Lambda}^{2(1+\gamma)}}
\operatorname{Li}_{2+\gamma}(z)-
\frac{L^{2(1+\gamma)}T^{1+\gamma}}{{\Lambda}^{2(1+\gamma)}}
\operatorname{Li}_{1+\gamma}(z)
\frac{\mu}{T} \\
&=\frac{\pi^{\gamma+1}T^{1+\gamma}}{{\Lambda}^{2(1+\gamma)}}\big[(2+\gamma)
\operatorname{Li}_{2+\gamma}(z)-\operatorname{Li}_{1+\gamma}(z)\frac{\mu}{T}\big].
\nonumber
\end{align}

The maximum at $\mu=0$ is
\begin{equation}\label{25}
S_{\mu=0}=\big(\frac{\pi}{\Lambda^2}\big)^{\gamma+1}(2+\gamma)\zeta(\gamma+2)T^{\gamma+1}.
\end{equation}
We are interested in $\gamma<0$.

Thus, we have two unknown constants, namely, $\Lambda$ and the value of the
entropy $S_{\mu=0}=\operatorname{const}$. We define these two constants from
the experimental value of the critical point of the liquid phase at the
negative pressure (see Sec.~6 below), namely, from the minimum point of the
pressure for a given simple liquid of the value $\gamma$ at this point and from
the temperature. This point is absent in the van der Waals model. This point is
present in our model of liquid phase.

For example, for water, we obtain $S_{\mu=0}=3.495$ and $\Lambda=3.74$.
However, the computation is carried out under the assumption that the Zeno-line
is a line segment, whereas this segment becomes curvilinear for water at low
temperatures.

According to the van der Waals conception, we normalize as follows:
\begin{equation}\label{26}
T^{\mathrm{red}}=\frac{T}{T_{\mathrm{cr}}}, \qquad
P^{\mathrm{red}}=\frac{P}{P_{\mathrm{cr}}}.
\end{equation}

\begin{figure}[h] %7
\begin{center}
\includegraphics[width=7cm]{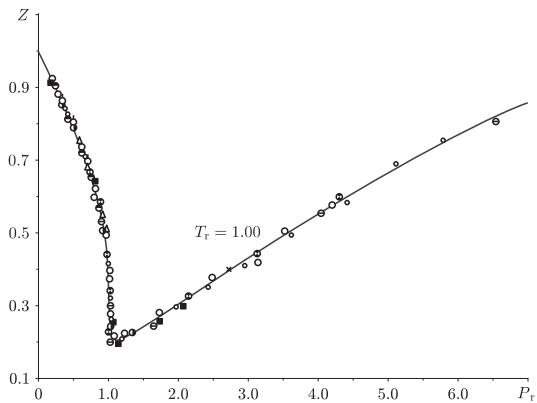}
\end{center}
\caption{Experimental graph for the different gases, including for methane, ethylene,
ethane, propane, $n$-butane, isopentane, $n$-heptane, nitrogen, carbon dioxide,
and water. Each gas is equipped with a particular symbol on the graph.}
\end{figure}

\begin{figure}[h] %8
\begin{center}
\includegraphics[width=7cm]{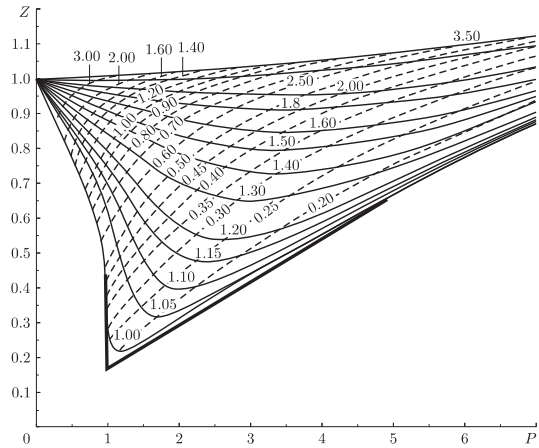}
\end{center}
\caption{The continuous lines are the experimental isotherms for $T\ge T_{\mathrm{cr}}$
for methane, the dotted lines are experimental isochors. The theoretical
critical isotherm coincides with the experimental isotherm up to $Z=0.29$ and
is continued by a straight line up to the point $P=1$, $Z=1/\rho_0=0.14$ (see
\thetag{27}--\thetag{29}). Further, at an acute angle, a tangent to the
experimental isotherm at a point of Zeno-line is drawn. The straight line from
the point $P=1,Z=0.14$ to the point of tangency is the critical isotherm of the
ideal liquid phase. It can be seen by comparing the figure with Fig~7, the
theoretical isotherm thus obtained corresponds to the isotherm of the ``law of
corresponding states'' for the gases indicated in Fig.~7.}
\end{figure}

Denote by $\rho_0$ the solution of the equation
\begin{equation}\label{27}
\frac{1}{T_B^{\mathrm{red}}}+\frac{\rho}{\rho_B}=1,
\end{equation}
where $T_B$ stands for the Boyle temperature and
\begin{equation}\label{28}
T_B^{\mathrm{red}}=\frac{T_B}{T_{\mathrm{cr}}}
\end{equation}
is a dimensionless quantity, $T^{\mathrm{red}}=\frac{T}{T_{\mathrm{cr}}}$. Then
\begin{equation}\label{29}
\frac{\rho_0}{\rho_B}=1-\frac{1}{T_B^{\mathrm{red}}}, \qquad
\rho_B^{\mathrm{red}}=\frac{\rho_B}{\rho_0}.
\end{equation}

Hence, the locus of the points of the quasi-ideal spinodals\footnote{That is,
of the endpoints of the metastable state of the liquid phase.} is given by the
formula
\begin{equation}\label{30}
P^{\mathrm{red}} =\frac{1}{\big(1-\frac{1}{T_B^{\mathrm{red}}}\big)}
T^{\mathrm{red}}\big(1-\frac{T^{\mathrm{red}}}{T_B^{\mathrm{red}}}\big)Z(\gamma),
\end{equation}
where $\rho_B$ and $T_B$ is the Boyle density and the Boyle temperature,
respectively.

Here, for $\gamma > 0$,
\begin{equation}\label{31}
Z_{\mathrm{cr}}^{\gamma(T^{\mathrm{red}})}=
\frac{\zeta(2+\gamma(T^{\mathrm{red}}))}{\zeta(1+\gamma(T^{\mathrm{red}}))}.
\end{equation}
Let us recall that $\gamma(T)$ can be calculated from the algebraic relation
$S_{\mu=0}=\mathrm{const}$.

If we use the Maxwell condition, see Sec~5 (if Sec~6 is taken into account),
which states that the transition from gas to liquid occurs for the same
chemical potential, pressure, and temperature, then we can construct the
so-called {\it binodal}. The binodal thus constructed coincides with the
experimental one, in contrast to the van der Waals binodal (Fig.~4\,(b)) and to
the binodal presented in Fig.~3.

\section{Wiener quantization of thermodynamics\\ and a jump of critical exponents}

The problem involving the so-called Maxwell rule for the transition
``gas--liquid,'' which is a natural complement of the new concept (of
phenomenological thermodynamics) constructed above, is solved by using the
tunnel (or Wiener) quantization introduced by the author already in his 1994 works;
see~\cite{26} and~\cite{27} and also~\cite{28},~\cite{29}, and~\cite{30}. We
repeat here this quantization at a heuristic level.

We can say that the quantization of thermodynamics is simply called for.
Indeed, we have the phase space\footnote{``A symplectic structure''.}\quad in which
the momenta are the extensive quantities~$V$ and~$-S$, and the corresponding
coordinates are~$P$ and~$T$. The usual quantization is of the form
\begin{equation}\label{32}
\widehat V=ih\frac\partial{\partial P}, \qquad -\widehat
S=ih\frac\partial{\partial T}.
\end{equation}
Just as in~\cite{31}, let us invoke an analogy between the Schr\"odinger
equation and the heat equation.

A. {\sl The Schr\"odinger equation\/} corresponding to a noninteracting
particle without an external field,
\begin{equation}\label{33}
-ih\frac{\partial\psi}{\partial t}=(ih\nabla)^2\psi.
\end{equation}
The change of variables
$$
\psi=e^{\frac ih\mathbb S}
$$
leads to the equation
$$
\frac{\partial\mathbb S}{\partial t}+(\nabla\mathbb S)^2+ ih\Delta\mathbb S=0.
$$
In this case, the quantization of the classical Hamilton--Jacobi equation
consists in the addition of the term $ih\Delta\mathbb S$.

B. {\sl The heat equation}
\begin{equation}\label{34}
-\nu\frac{\partial u}{\partial t}=(\nu\nabla)^2u,
\end{equation}
$\nu$ is the kinematic viscosity. The change of variables
$$
u=e^{-\frac{\mathbb S}\nu}
$$
leads to the equation
\begin{equation}\label{35}
\frac{\partial\mathbb S}{\partial t}+(\nabla\mathbb S)^2+ \nu\Delta\mathbb S=0.
\end{equation}
The derivatives of this equation with respect to the coordinates are called the
{\it Burgers equations}. In this case, the {\it  Wiener quantization\/} consists
in the addition of the viscous term.

\begin{remark} \rm
In this special case, the Wiener quantization coincides with
the Euclidean quantization well known in field theory. In the general case, it
corresponds to the passage from the Feynman path integral to the Wiener path
integral and is in essence presented in detail for physicists in the
book~\cite{32}.
\end{remark}

In the Burgers equation, for $p=\partial\mathbb S/\partial x$, a shock wave
occurs as $\nu\to0$, i.e., a discontinuity of the $\theta$-function type,
whereas, in thermodynamics, we have a jump of the $\theta$-function type for
the transition ``gas--liquid.'' For the Burgers equation, the rule of ``equal
areas'' arises. For the ``gas--liquid'' transition, the Maxwell rule of equal
areas arises. In the heat equation, the  Wiener quantization of energy yields
$\nu D_t$, where $D_t$ stands for the Heaviside operator,
$D_t=\frac\partial{\partial t}$. In thermodynamics, the thermodynamical
potential, the Gibbs energy, is equal to $\mu N$, where $\mu$ stands for the
chemical potential and $N$ for the conjugate extensive quantity, the number of
particles. Hence,
$$
\widehat N=\nu\frac\partial{\partial\mu},
$$
and the role of time\footnote{Cf.~Matsubara Green function,
where the role of  the imaginary time is played by the parameter $\beta=1/T$~\cite{65}.}
is played by $\log-\mu$, because, under this quantization,
the operator $\nu\mu(\partial/{\partial\mu})$ corresponds to the Gibbs energy.

For us, the one-dimensional case $p_1=V$ and $q_1=P$ is of importance. In the
general case, the focal point (the point of inflection~\cite{34},~\cite{35}) is
of the form $q\sim p^3$, i.e.,
\begin{equation}\label{36}
P_{\mathrm{cr}}\sim(V-V_{\mathrm{cr}})^3=
\bigg(\frac{\rho_{\mathrm{cr}}-\rho}{\rho\rho_{\mathrm{cr}}}R\bigg)^3,
\end{equation}
which corresponds to the classical critical index (the exponent) equal to
three. The asymptotic solution (as $\nu\to0$) of~\eqref{35} corresponding to
this point is expressed by the Weber function. The critical point itself
corresponds to the point $x=0$.

The Weber function is of the form
\begin{equation}\label{37}
u(x)\cong\frac1{\sqrt\nu}\int_0^\infty e^{-\frac{x\xi-\frac{\xi^4}4}\nu}\,d\xi.
\end{equation}

If we set $x=0$, then the change of variables
$$
\xi=\root4\of\nu y^{1/4}
$$
gives a singularity as $\nu\to0$ of the order of $\nu^{-1/4}$. What does it
mean, from the classical point of view and classical measurement, for the
case in which the condition referred to in~\cite{20} as the ``semiclassical
condition'' holds (this means that we are outside the focal point)? For the
Laplace transform, this means that we are in a domain in which the Laplace
asymptotic method is applicable, i.e., in a domain for which
\begin{equation}\label{38}
\Psi(x)=\frac1{\sqrt\nu}\int_0^\infty e^{-\frac{px-\widetilde S(p)}\nu}\,dp,
\qquad \lim_{p\to0}\frac{\widetilde S(p)}{p^4}<\infty, \qquad \widetilde
S^{(4)}(p)|_{p=0}\ne0.
\end{equation}

The solution $p_\nu(x)$ of the Burgers equation can be evaluated by the formula
\begin{equation}\label{39}
p_\nu(x)=\nu\frac{\partial\log u(x)}{\partial
x}=\frac{\int_0^\infty\exp\{\frac{-x\xi-\widetilde
S(\xi)}\nu\}\xi\,d\xi}{\int_0^\infty\exp\{\frac{-x\xi-\widetilde
S(\xi)}\nu\}d\xi}.
\end{equation}
After the substitution
$$
\frac\xi{\root4\of\nu}=y,
$$ as $x\to0$ we obtain
\begin{equation}\label{40}
p_\nu(x)\to_{x\to0}{\root4\of\nu}\cdot \text{const}.
\end{equation}
In our case, the momentum $p_\nu(x)$ is the volume~$V$.

If the solution of the relation
$$
x={\partial\widetilde S}/{\partial p}
$$
is nondegenerate, i.e.,
$$
\frac{\partial^2\widetilde S}{\partial p^2}\ne0
$$
at the point
$$
\frac{\partial\widetilde S}{\partial p}=x,
$$
then in this case, the reduced integral~\eqref{39} is bounded as $\nu\to0$.
For this integral to have
a singularity of order $\nu^{1/4}$, we must apply to this integral the
fractional derivative $D^{-1/4}$ with respect to $x$. The value of $D^{-1/4}$
at the function equal to one, $D^{-1/4}1$, gives approximately $x^{1/4}$.

In our case, the pressure $P$ plays the role of $x$, and the volume $V$ plays
the role of momentum $p$. Therefore, $V\sim P^{1/4},$ i.e.,
$$
P_{\mathrm{cr}}\sim(V-V_{\mathrm{cr}})^4\sim
\big(\frac{\rho-\rho_{\mathrm{cr}}}{\rho\rho_{\mathrm{cr}}}R\big)^4.
$$

Following M.~Green~\cite{36}, D.~Yu.~Ivanov, a deep experimenter, poses the
following question: Why are the deviations from the classical theory in the
critical opalescence observed within the limits of hundredths of a degree
from the critical points, whereas the deviations in thermodynamical properties
show~\cite{7} a nonclassical behavior at a much larger distance from the
critical point?. Professor Ivanov claims that rather many questions of this
kind have been accumulated (see, for example,~\cite{8}), and all these
questions mainly deal with the behavior of practical systems. The point is
that, from the point of view of the developed theory of critical
indices~\cite{6}, there must be a drastic passage to the classical indices
outside a neighborhood of the critical point.

To make Ivanov's question understandable for persons who are not experts in
critical points, we paraphrase the question for the case of geometric optics,
when the sun rays are collected by a magnifying glass to a focus. If we had
created the special construction for the vicinity of the focus in which the paper
smoulders, then the experimenter could ask why the experiment gives a smooth
picture of transition in the double logarithmic coordinates and the indices are
preserved far away from the smouldering small vicinity of the focus. In the present
case, the smouldering paper can be compared with the small area of opalescence
(drastic fluctuations near the critical point) for which a separate theory was
constructed. At the same time, the special function defining the point within
the wave theory (like the Weber function) can be continued quite smoothly to a
much wider domain in which the paper does not smoulder. In the opinion of
Ivanov, this fact is much more important than the fact that Wilson's theory
gives the index 4.82 rather than 4.3, which is given by modern experiments.

Can the experimental index 4.3 be explained in principle in the framework of
the conception presented by the author?

Whereas in classical mechanics there is no dependence on the Planck constant
$\hbar$, in classical thermodynamics there is a slow dependence of the
viscosity on $T$ and $\rho$, and thus {\it vice versa\/} as well. In our
picture, the ``stretching'' of $P_{\mathrm{cr}}$ and $T_{\mathrm{cr}}$ in the
experiment for real gases (Figs.~7 and~8) is greater than in the van der Waals
model, which enables us to introduce in the ``stretching'' the parameter
$\nu^\varepsilon$ (i.e., $P=\nu^\varepsilon V^3$) and obtain the index
$\delta=4.3$ for $\varepsilon=0.07$. Thus, in principle, the answer can be
``yes.''

It follows from what was said above that new critical indices arise only due to
quantization of the conjugate pairs $\{P,V\}$ and $\{T,S\}$. Thus, the
relationships between intensive quantities can be taken classical, because
everything is carried out under the assumption of infinitely small viscosity.
In this case, one can also pass to another coordinates, to the pressure and
density.

Write, as usual,
$$
p=\frac{P-P_{\mathrm{cr}}}{P_{\mathrm{cr}}}, \qquad
\theta=\frac{T-T_{\mathrm{cr}}}{T_{\mathrm{cr}}}, \qquad
v=\frac{\rho-\rho_{\mathrm{cr}}}{\rho_{\mathrm{cr}}}.
$$
In the classical case, we have (\cite{37}, p.~344)
$$
p\sim v^3, \qquad v\sim\theta^{1/2}, \qquad \theta\sim v^2.
$$
In the classical case, we obtain
$$
p\sim\theta^{3/2}.
$$

In the tunnel quantum case, we obtain $\beta=0.375$ (cf.~\cite{37}, p.~356) in
the limit as $\nu\to0$, and thus not precisely (the stretching is not taken
into account). This is obtained for the van der Waals model quantized in the
tunnel way.

An important specific feature of  Wiener quantization is that, at the expense of
viscosity, it smoothes the acute angle on the graphs~8 and~9, and thus removes
the discrepancy between the theory and the experiment.

The heuristic role of viscosity when penetrating through the barrier is
described in Sec.~2. As is well known, an exponential penetration through the
barrier occurs in the standard quantization as well. However, in the Wiener
quantization, we restrict ourselves to a power-law approximation (see Sec.~6 below).

If $\nu>0$, then an uncertainty principle arises. Let us proceed with the
consideration of this principle.

Thus, in the case of Wiener quantization,
the main operators are not the momentum operator
$ih \frac{\partial}{\partial x}$ and the operator of multiplication by $x$,
but the Heaviside operator $\nu D=\nu\frac{\partial}{\partial x}$
and the operator of multiplication by $x$.
The constant $\nu$ was determined in~\cite{38} as viscosity.
In contrast to the Plank constant, the constant $\nu$ can smoothly depend on
the variables of the system.
The main distinction of the tunnel quantization from
the Euclidean and usual quantization is that it is considered up to $O(\nu^k)$,
where $k$ is any fixed number, i.e., it is factorized in $O(\nu^k)$.

But first it is necessary to define the space where these operators act.

As is well known, the Heaviside operator is related to the two-sided Laplace
transform. This was shown already by van der Pol and Bremmer in~\cite{39}.
Introduce a family of functions $\varphi(p)$ to which we shall apply the
two-sided Laplace transform, namely,
\begin{equation}\label{41}
\varphi(p)=\int_0^\infty e^{-p^2\xi}\Xi(\xi)\,d\xi,
\end{equation}
and thus these functions themselves are one-sided Laplace transforms,
$$
F_\lambda\Xi(\xi)=\int_0^\infty e^{-\lambda\xi}\Xi(\xi)\,d\xi,
$$
for $\lambda=p^2$. Denote by $F_\lambda^\pm$ the two-sided Laplace transform,
$$
F_\lambda^\pm\varphi(p)=\int_{-\infty}^\infty e^{-\lambda p}\varphi(p)\, dp.
$$

If the functions $\Xi(\xi)$ are compactly supported and infinitely
differentiable, then the closure of the operator~$\nu D$ with respect to this
domain can be carried out in the Bergman space. Then the functions
$\Psi(x)=F_x^\pm\varphi(p)$ here become analogous to the $\Psi$-functions in
the Schr\"odinger quantization. Moreover, $\Psi^*(x)=\Psi(x)$, because these
functions are real-valued.

Let us note first of all that the squared function or, equivalently, the
squared dispersion $\Delta\widehat f$ of the operator $\widehat f$ is
$$
(\Delta\widehat f)^2=\big|(\widehat f-\overline f)^2\big|,
$$
and, since $\widehat f$ is not selfadjoint, the function $(\widehat f-\overline
f)^2$ need not be positive, and hence one must pass to its absolute value.
Therefore, the corresponding theorem for generic operators fails to hold in
general. However, for the operators $\nu D$ and $x$ on a reduced function
space, we obtain $ |\Delta\nu D||\Delta x|\ge\nu/2. $ It can readily be seen
that Weyl's proof (which is presented in the comments to \S16 of Chap.~II
in~\cite{20}) can easily be transferred to the operators $\nu D$ and $x$ in the
above function space.

Let us note some consequences of tunnel (or Wiener) quantization for the ``quantum'' Bose
gas.

A specific feature of photon gas, which is mentioned in~\cite{9}, \S\S~62,63,
is that the number of particles in this gas, $N$, is a variable quantity
(rather than a given constant, which is the case for ordinary gas).

Thus, since the number of particles $N$ in thermodynamics is conjugate to the
chemical potential, it follows that, if the number of particles is undefined,
then the chemical potential can be given precisely, $\mu=0$, under the
assumption that $\mu$ and $N$ are tunnel quantized and the uncertainty
principle holds.

A contradiction between the conception of the author and the conception of
physicists going back to Einstein is also removed. In the case of a gas for
which $N$ is fixed, we have
\begin{equation}\label{42}
\sum^\infty_{i=0}N_i=N
\end{equation}
according to the relation in~\cite{9}, and the chemical potential $\mu$ can be
a small positive quantity. This is obvious, because $N_i\le N$; however, this
contradicts Einstein's original conception claiming that $\mu\le0$. This
contradiction is removed if the relationship of the uncertainty principle holds
for $\mu$ and $N$, because, if $\mu=0$, then $N$ can take infinite values as
well, and therefore the case $\mu>\nobreak0$ is impossible.

Thus, it can be said that both the scaling hypothesis and the hypothesis of
tunnel quantization do not agree near the critical point with the old
thermodynamical conception of the four potentials. However, the hypothesis of
tunnel quantization does not contradict the conception of the four potentials,
namely, the hypothesis complements the conception, and this works not only near
the critical point but also on the entire domain ``gas--liquid'' by agreeing
with the Maxwell rule and by removing logical discrepancies in Bose gas
theory.

We have explained Wiener quantization, which enabled us to remove some
problems. Let us now also explain the second quantization. The principal
element in Fock's approach is the indistinguishability of particles. In our
theory, it follows from the original axiom. Although there are no natural
Hilbert spaces here, in contrast to quantum mechanics, we can still obtain
correct distinguished representations and limits as $h\to0$ (see~\cite{40},
Chap.~1, Appendix~1.A) and then, in view of the indistinguishability of
particles, perform the second quantization of classical theory by introducing
the creation and annihilation operators. Certainly, this is possible only under
the condition of the {\it principle of indistinguishability of particles in our
measurements}, which follows from the main axiom, rather than from the
``identity principle'' introduced in~\cite{20}.

In classical mechanics, such operators were introduced in~\cite{41} and
\cite{42} on the basis of the Sch\"onberg concept (see~\cite{43}
and~\cite{44}).

Thus, the contemporary derivation of the Vlasov equation is obtained by
applying the method of second quantization for classical particles~\cite{40}.
In this case, as $N\to\infty$, one obtains a system for which the creation and
annihilation operators asymptotically commute,
\begin{align}\label{43}
\dot{u}(p,q,t)
&=  \biggl(\frac{\partial U}{\partial q}
\frac{\partial}{\partial p}-p\frac{\partial}{\partial q}\biggr) u(p,q,t)
\nonumber\\
&\qquad\int dp'dq'v(p',q',t) \biggl(\frac{\partial V(q,q')}{\partial q}
\frac{\partial}{\partial p}+
\frac{\partial V(q,q')}{\partial q'}
\frac{\partial}{\partial p'}\biggr) u(p',q',t) u(p,q,t),
\nonumber\\
\dot {v}(p,q,t)
&= \biggl(\frac{\partial U}{\partial q}\frac{\partial}{\partial p}
-p\frac{\partial}{\partial q}\biggr) v(p,q,t)
\\
&\quad+\int dp'dq'u(p',q',t) \biggl(\frac{\partial V(q,q')}{\partial
q}\frac{\partial}{\partial p}+ \frac{\partial V(q,q')}{\partial q'}
\frac{\partial}{\partial p'}\biggr) v(p',q',t) v(p,q,t),
\nonumber
\end{align}
where $U(q_i)$ is an external field and $V(q_i,q_j)$ is the pairwise
interaction.

If one replaces $u$ and $v$ by the operators of creation and annihilation
$\widehat{u}$ and $\widehat{v}$ in the Fock space, then, after this change,
system~\eqref{43} becomes equivalent to the $N$-particle problem for the Newton
system.

However, according to the mathematical proof, this can happen only for the case
in which the classical particles are indistinguishable (from the point of view
of the main axiom).

Only in this case the projection from the Fock space to the $3N$-dimensional
space of $N$ particles gives {\it precisely\/} the system of Newton equations.

Note that the substitution
\begin{equation}\label{44}
u(p,q,t)=\sqrt{\rho(p,q,t)}e^{i\pi(p,q,t)}, \qquad
v(p,q,t)=\sqrt{\rho(p,q,t)}e^{-i\pi(p,q,t)},
\end{equation}
reduces system \eqref{43} to the form
\begin{align}\label{45}
\dot {\rho}(p,q,t)
&=\biggl(\frac{\partial W^t}{\partial q}
\frac{\partial}{\partial p}-p\frac{\partial}{\partial q}\biggr) \rho(p,q,t),
\nonumber\\
\dot {\pi}(p,q,t)
&=\biggl(\frac{\partial W^t}{\partial
q}\frac{\partial}{\partial p}-p\frac{\partial}{\partial q}\biggr)
\pi(p,q,t)+\int dp'dq'\frac{\partial V(q,q')}{\partial q'} \frac{\partial
\pi(p',q',t)}{\partial p'} \rho(p',q',t);
\end{align}
we have here
$$
W^t(q)=U(q)+\int dq'V(q,q') \rho(p',q',t) dp'dq'.
$$

The first equation of system~\thetag{45} is the {{\it Vlasov}} equation
(see~\cite{56}), where $\rho$ stands for the distribution function and $W^t(q)$
for the dressed potential (see Sec.~2, formulas~\eqref{5}ff.); the other
equation is linear, and its meaning is discussed in~\cite{59}.

Note further a tunnel-quantum jump of the index at the points of the spinodal
of the liquid phase. The classical index of the spinodal is equal to 2, namely,
$P\sim V^2$, similarly to turning points in quantum mechanics. The Airy
function corresponds to it. Similarly to~\eqref{36}--\eqref{38}, we obtain
\begin{equation}\label{46}
\Psi(x)= \frac {1}{\sqrt{\nu}} \int_0^\infty
e^{-\frac{px-\widetilde{S}(p)}{\nu}} \, dp, \qquad \lim_{p\to 0}
\frac{\widetilde{S}(p)}{p^3} < \infty, \qquad
\widetilde{S}^{(3)}(p)|_{p=0}\neq0.
\end{equation}

The solution $p_\nu(x)$ of the Burgers equation can be evaluated by the
formula~\eqref{37}. As $x\to 0$, after the change
$\frac{\xi}{\root3\of{\nu}}=y$, we obtain
\begin{equation}\label{47}
p_{\nu}(x) \rightarrow_{x\to 0}  {\root3\of{\nu}}\cdot \text{const}.
\end{equation}
In our case, the momentum $p_\nu(x)$ is the volume~$V$. Hence, similarly to the
consideration between the formulas~\eqref{39} and \eqref{40}, we obtain $P\sim
V^3$, and the index at the points of the spinodal becomes equal to three.

\begin{remark} \rm
It is possible that an experimenter, when considering the
approaching of the critical isotherm for $T>T_{\mathrm{cr}}$ to the critical
point $\mu =0$, moves (due to the indeterminacy principle) towards increasing
values of~$N$, and hence towards increasing density, and arrives at the
spinodal of the liquid phase. This effect is similar to the accumulation of the
wave crest which overturns afterwards (a part of the particles outruns the
point of creation of the shock wave). In this case, the critical index 4.3
passes to the index 3 of the spinodal (and this index occasionally coincides
with the classical index of the critical point). This passage, which is
described by the Vlasov equation, was experimentally noticed in~\cite{8} and in
other works. Therefore, the experiments of Ivanov~\cite{8} and
Wagner~(\cite{57},~\cite{58}), where the modifications of the critical index
$\delta$ from 4.3 to 3 were obtained when approaching the critical point, do
not contradict our conception.
\end{remark}

For the creation of
dimers, the author of the present paper used the creation and annihilation
operators for pairs of particles and referred to it as the ultrasecond
quantization~\cite{30}. Experimenters do not distinguish between dimers either,
counting only their number (for example, as showed by Calo~\cite{45}, the
presence of 5--7\% dimers leads to the appearance of a cluster cascade).

Thus, we discover new relations, namely an extension of the program ``partitio
numerorum'' in number theory, from the point of view of the notion of Hartley
entropy, and indicate possible generalizations of quantization, which lead to
an extension of the Heisenberg indeterminacy principle~\cite{25}.

Ultrasecond quantization leads to thermodynamics in nanocapillaries and
enabled us to obtain the superfluidity of liquids in
nanotubes~\cite{46}--\cite{48}, which was confirmed in experiments
(see~\cite{49}--\cite{51}).

\section{Negative pressure and a new critical point of possible transition
from liquid to ``foam''}

First, we use the Wiener quantization of thermodynamics
to explain why, in Sec.~5,
the model of incompressible liquid was used to derive the condition
$S|_{\mu=0}=\mathrm{const}$,
where the constant is independent of the temperature.

Because of the Bachinskii relation on the Zeno line,
the incompressible liquid model leads to a rigid relation
between the density (concentration)~$\rho$ and the temperature.
Since the value of $N$ is undefined for $\mu=0$
(i.e., the concentration $\rho$ is undefined),
the temperature is also undefined, and hence,
according to the indeterminancy principle,
the entropy takes a constant value and can be determined.
(In addition, we note that, for $\mu=0$, the activity is equal to~$1$
at any (undefined) temperature).
This precisely means that $S|_{\mu=0}=\mathrm{const}$.
As we shall see below, the value of this constant is uniquely determined
by the new critical point at which the liquid passes into ``foam''
(dispersion state).
The value of this point also determines the constant $\Lambda$
in the definition of $\Omega$, and hence, after calculating
$\varphi_{\gamma}(V/V_{\mathrm{cr}})$,
it is possible to obtain the complete description
of the distribution function for the gas branch in thermodynamics.
We note that the indeterminancy at the spinodal point
agrees with the results of experiments
on the absolute instability of the spinodal point,
which were performed by academician Skripov and his school.

Now we pass to negative $Z$.

As is known, the Bose--Einstein distribution is obtained as the sum of terms of
an infinitely decreasing geometric progression. If the progression is bounded
by the number $N$, then the potential is of the form
\begin{equation}\label{48}
\Omega=-T\sum_k\log\Big(\frac{1
-\exp\frac{\mu-\varepsilon_k}TN}{1-\exp\frac{\mu-\varepsilon_k}T}\Big).
\end{equation}

What is the relationship\footnote{The physicists who are not interested
in Euler--Maclaurin-type bounds used to pass from sums to integrals
can skip the following scheme of the proof of these bounds.}
between $E_i$ and $\gamma$?

1) Ultrarelativistic case. $E=cp$. Here
\begin{equation}\label{49}
E_{i+1}-E_i=\int_{E_i}^{E_{i+1}} cp\, p^2\, dp= \frac 14\bigg( p^4(E_{i+1}) -
p^4(E_{i})\bigg) \sim \frac 34 cp^3(E_i^{i+1})=\frac{3}{c^2}
(E_i^{i+1})^3.
\end{equation}

\medskip

2) Nonrelativistic case. $E=\frac{p^2}{2m}.$ Here
\begin{align}\label{50}
E_{i+1}-E_i&=\int_{E_i}^{E_{i+1}}\frac{p^2}{2m} p^2\, dp=
\frac{1}{2m} \bigg(\frac{p^5}{5} (E_{i+1}) -\frac{p^5}{5} (E_{i})\bigg)
\nonumber\\
&= \frac{1}{2m}\bigg(\frac{(\sqrt{2m E_{i+1}})^5}{5} -\frac{(\sqrt{2m
E_{i}})^5}{5}\bigg)\cong \mathrm{const}  (E_i^{i+1})^{3/2}.
\end{align}

\medskip

3) Consideration of the degrees of freedom. $E=\frac{p^{2+\sigma}}{2mp_0}.$
Here
\begin{equation}\label{51}
E_{i+1}-E_i=\int_{E_i}^{E_{i+1}} \frac{p^{2+\sigma}}{2m} p^2\, dp \cong
\mathrm{const}  (E_i^{i+1})^{(4+\sigma)/(2+\sigma)}.
\end{equation}

By~\eqref{15}, $ \gamma=({1-\sigma})/({2+\sigma}), $ and thus $\gamma <0$ for
$\sigma >1$.

As was proved in~\cite{17}--\cite{18}, passing to the limit in the
Euler--Maclaurin formula, we obtain
\begin{equation}\label{52}
N=\frac{1}{(\gamma+1)\Gamma(\gamma+1)} \int_0^\infty
\left\{\frac{1}{e^{b\xi}-1}-\frac{N}{e^{N b\xi}-1}
\right\}\,d\xi^\alpha.
\end{equation}
In particular, for $\gamma = -1/2$,
\begin{equation}\label{53}
N=\frac{1}{\Gamma(3/2)} \int_0^\infty
\left\{\frac{1}{e^{b\xi^2}-1}-\frac{N}{e^{N b\xi^2}-1} \right\}\,d\xi.
\end{equation}
The absolute value of the derivative of the integrand can readily be estimated
by using the identities presented below. By the Euler--Maclaurin bounds, this
shows that one can pass from the sums of the form~\eqref{41} to the
corresponding integrals with the accuracy needed here.

Hence, denoting $N_{\operatorname{cr}}=k_0$, we obtain the following formula
for the integral at $\mu=0$:
\begin{equation}\label{54}
\mathcal{E}= \frac{1}{\alpha\Gamma(\gamma+2)}\int \frac{\xi
\,d\xi^\alpha}{e^{b\xi}-1} = \frac{1}{b^{1+\alpha}}\int_0^\infty\frac{\eta
d\eta^\alpha}{e^\eta-1},
\end{equation}
where $\alpha=\gamma+1$. This implies that
\begin{equation}\label{55}
b=\frac{1}{\mathcal{E}^{1/(1+\alpha)}}
\left(\frac{1}{\alpha\Gamma(\gamma+2)}\int_0^\infty\frac{\xi
\,d\xi^\alpha}{e^\xi-1}\right)^{1/(1+\alpha)}.
\end{equation}

We obtain
\begin{align}\label{56}
&\int_0^\infty
\left\{\frac{1}{e^{b\xi}-1}-\frac{k_0}{e^{k_0b\xi}-1}
\right\}\,d\xi^\alpha =\frac{1}{b^\alpha}\int_0^\infty\left(\frac{1}{e^\xi-1}
-\frac{1}{\xi}\right)\,d\xi^\alpha
\nonumber\\
&\qquad+\frac{1}{b^\alpha}\int_0^\infty
\left( \frac{1}{\xi}- \frac{1}{\xi(1+(k_0/2)\xi)}\right)\,d\xi^\alpha  - \frac{
k_0^{1-\alpha}}{b^\alpha}\int_0^\infty\left\{ \frac{k_0^\alpha}{e^{k_0\xi}-1}
-\frac{k_0^\alpha}{k_0\xi(1+(k_0/2)\xi)}\right\}\,d\xi^\alpha.
\end{align}
Write
$$
c=\int_0^\infty\left(\frac1\xi-\frac1{e^\xi-1}\right) \xi^\gamma\,d\xi.
$$
After the change $k_0\xi=\eta$, we obtain
\begin{align}\label{57}
&\frac{ k_0^{1-\alpha}}{b^\alpha}\int_0^\infty\left\{
\frac{k_0^\alpha}{e^\eta-1}-\frac{k_0^\alpha}{\eta(1+\eta/2)}
\right\}\,d\xi^\alpha =\frac{k_0^{1-\alpha}}{b^\alpha}\int_0^\infty
\left\{\frac1{e^\eta-1}-\frac1{\eta(1+\eta/2)}\right\}d\eta^\alpha
\nonumber\\
&\qquad= \frac{k_0^{1-\alpha}}{b^\alpha}
\left\{\int_0^\infty\left(\frac1{e^\eta-1}-\frac1\eta\right)+
\int_0^\infty\frac{d\eta^\alpha}{2(1+\frac\eta2)}\right\}=-c
\frac{k_0^{1-\alpha}}{b^\alpha}+ c_1\frac{k_0^{1-\alpha}}{b^\alpha}\,.
\end{align}
Since
$\frac1{\eta(1+\eta/2)}=\frac1\eta-\frac1{2(1+\eta/2)},$
after denoting
$c_1=\int_0^\infty\frac{d\eta^\alpha}{2(1+\frac\eta2)}\,,$
we can write
\begin{equation}\label{58}
\int_0^\infty\left(\frac1\xi -\frac1{\xi(1+\frac{k_0}2\xi)}\right)\,d\xi^\alpha
=\frac{k_0}2\int_0^\infty \frac{\,d\xi^\alpha}{1+\frac{k_0}2\xi}=
\left(\frac{k_0}2\right)^{1-\alpha}\int_0^\infty
\frac{d\eta^\alpha}{1+\eta}=c_1\left(\frac{k_0}2\right)^{1-\alpha}.
\end{equation}
Hence,
\begin{align}\label{59}
&-\frac1{b^\alpha}c_1+\frac1{b^\alpha}c\left(\frac{k_0}2\right)^{1-\alpha}
-\frac{k_0^{1-\alpha}}{b^\alpha}\int_0^\infty
\left\{\frac1{e^\eta-1}-\frac1{\eta(1-\frac\eta2) }\right\}d\eta^\alpha
-\frac12\int\frac{d\eta^\alpha}{1+\frac\eta2}\cdot
\frac{k_0^{1-\alpha}}{b^\alpha}
\nonumber\\
&\quad=-\frac1{b^\alpha}c+\frac{k_0^{1-\alpha}}{b^\alpha}c.
\end{align}

Since $k_0$ is the number of particles, $b=1/T$, and $\alpha=1+\gamma$, it
follows that $k_0b^\alpha$ for $\gamma>0$ is the value of the Riemann zeta
function, $\zeta(1+\gamma)$. Therefore, $k_0^{\gamma+1}$ increases for
$\gamma<1$, and the first term of the right-hand side of equation~\eqref{59}
can be neglected. Introducing the function
\begin{equation}\label{60}
\mathcal M(\gamma+1)=
\big({c(\gamma)}/{\Gamma(\gamma+1)}\big)^{1/({1+\gamma})},
\end{equation}
we see that the compressibility factor $ Z_\gamma=-\zeta(\gamma+2)/\mathcal
M(\gamma+1) $ is subjected to a flexion\footnote{Since the ``Young moduli'' for
the compression and extension are distinct, a flexion of the spinodal occurs.}
\ from $\gamma>0$ to
$\gamma<0$.

In this case, we obtain another critical point, which fully corresponds to the
physical meaning (see~~\cite{53}).

Thus, if the compressibility factor is negative, then we divide $\mathcal{E}/N$ by
$T^{\gamma+1}$ with $\gamma <0$ rather than by $T$, because
$$
\frac{\mathcal{E}}{N}\Big|_{\mu=0}
=\frac{\zeta(\gamma+2) T^{\gamma+2}_{\text{red}}} {\mathcal{M}(\gamma+1)T_{\text{red}}}
= T^{\gamma+1}_{\text{red}}\frac{\zeta(\gamma+2)}{\mathcal{M}(\gamma+1)},
$$
i.e., the energy evaluated for a
single particle at $\mu=0$ (at the ``degeneration'' point), for $P<0$, is
proportional to $T^{\gamma+1}_{\mathrm{red}}$, i.e., to the temperature taken
to a power with an exponent less than one. For $P<0$, the compressibility
factor becomes a dimensional quantity; however, this is always considered in
this very way on curves in the space$\{Z,P\}$ when using the van der Waals
normalization~\thetag{26}.

We write
$$
F(\xi)=\left(\frac 1 \xi -\frac{1}{e^{\xi}-1}\right).
$$
For $\gamma <0$ and $\mu=0$ we have
\begin{equation}\label{61}
N_c= \sum_{j=1}^\infty \frac{j^\gamma}{e^{bj}-1}=
\sum_{j=1}^\infty j^\gamma\frac{1}{bj}-
\sum_{j=1}^\infty j^\gamma F(bj).
\end{equation}
Since the function $f(x)=x^\gamma F(bx)$ decreases monotonically, we have
\begin{equation}\label{62}
\sum_{j=1}^\infty j^\gamma F(bj)=
\sum_{j=1}^\infty  f(j) \leq
\int_0^\infty f(x)\, dx=
\int_0^\infty x^\gamma F(bx)\, dx=
b^{-\gamma-1} \int_0^\infty x^\gamma F(x)\, dx
\end{equation}
(the Nazaikinskii inequality).

Thus, the spinodal point $\gamma(T)$ is determined by the relation
$$
N=b^{-1}\zeta(1-\gamma)+O(b^{-1-\gamma}), \qquad \gamma <0,
\qquad b=\frac {1}{T_\mathrm{r}},
$$
where $\zeta$ is the Riemann zeta function.

For the critical isotherm we have $T_\mathrm{r}=1$.
Hence $\gamma(T)=-\gamma_0$.
For the point $T_\mathrm{r}$ and the isochor
$T_\mathrm{r}^{\gamma_0+1}\zeta(\gamma_0+1)$
we obtain the relation for $\gamma(T_\mathrm{r})$
from the equation
\begin{equation}\label{63}
T_\mathrm{r}^{\gamma_0+1}\zeta(\gamma_0+1)= T_\mathrm{r}\zeta(1+|\gamma(T_\mathrm{r})|).
\end{equation}
The less $T_\mathrm{r}$, the greater $|\gamma(T)|$.
But $\gamma(T)$ cannot exceed~$1$,
because the pressure and $Z$ become $-\infty$ at this point.

The negative pressure at the point of the spinodal of liquid phase
(for $\tilde{\mu}=0$) is equal to
$-T^{2+\gamma(T)}\zeta(2+\gamma(T_\mathrm{r}))$,
where the value of $\gamma(T)$ obtained above is negative.
The condition $\gamma(T) \leq 1$ bounds the transition region
for the temperature
\begin{equation}\label{64}
T_\mathrm{r}^{\gamma_0} \leq  \frac{\zeta(2)}{\zeta(\gamma_0+1)}=
\frac{\pi^2}{6 \zeta(\gamma_0+1)}.
\end{equation}

For a model we can consider thick rubber with many small slots
(see Appendix in~[33), which is almost incompressible in compression,
and because of the holes--slots,
has a small Young modulus in tension.
It is natural to consider tension of a liquid
as the appearance of nanoholds, i.e., of negative energy $PV$,
which can be treated as negative pressure
(pressure of the holes).

Let us now present a graph Fig.~9 for negative pressure for the Lennard--Jones
potential, where the new critical point is obtained by using a computer
experiment\footnote{The absolute zero of temperature is inaccessible. This is
visually seen in the logarithmic scale of temperatures $\log T_{\mathrm{red}}$,
where the absolute zero corresponds to~$-\infty$.}.

\begin{figure}[h] %9
\begin{center}
\includegraphics[width=10cm]{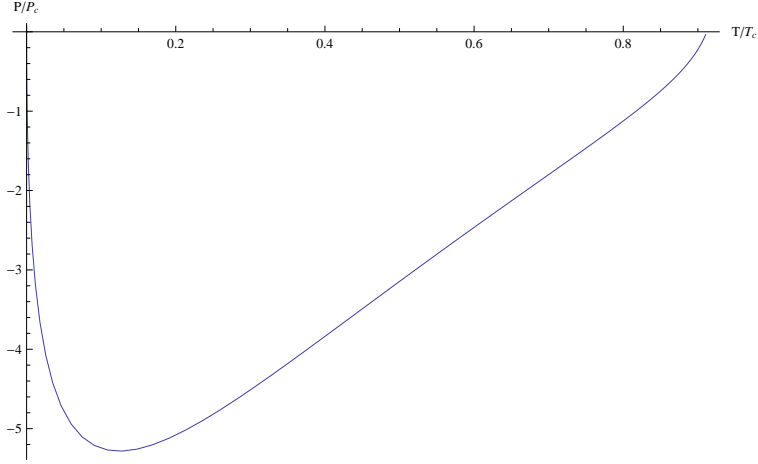}
\end{center}
\caption{The spinodal in the coordinates given by the temperature $T_{\mathrm{red}}$
and the negative pressure $P_{\mathrm{red}}$.}
\end{figure}

\section{On Homogeneous Mixtures of Gases}

When considering a gas mixture, we would like to attract the attention at the
following fundamental point. As is well known, in statistical physics and
thermodynamics, the energy is sometimes connected with the number of degrees of
freedom and the temperature; for example, this is the case in the
equidistribution law. It turns out in this case that the energy depends on the
temperature and on the number of degrees of freedom and does not depend on the
mass. The sequential usage of this conception gave us a continuous parameter
$\gamma$ related to the fractal dimension in the momentum space. Continuing the
use of this conception in the case of a mixture of pure gases, we should speak
of concentration rather than on density, i.e., we should neglect the
masses of miscible pure gases.

The sequential application of number theory in thermodynamics, i.e., the
consideration of the main axiom for the gas mixture (which was in fact made by
experimenters, at least in the case of air (see~\cite{55})), leads to the
formulas presented below.

If two values $\mathcal E_1$ and $\mathcal E_2$ expanded into sums of $N_1$ and
$N_2$ summands, respectively, correspond to the fractional
dimensions~$\gamma_1$ and~$\gamma_2$, respectively, and the values of the pairs
$\{\mathcal E_1,N_1\}$ and $\{\mathcal E_2,N_2\}$ are on the ``verge of
degeneration,'' i.e., adding an excessive number to $N_1$ and to $N_2$ leads to
the ``appearance of the Bose condensate,'' then, for the sum $\mathcal
E_1+\mathcal E_2$ and for $N_1+N_2$, any adding of an excessive number to
$N_1+N_2$ also leads to the ``appearance of the Bose condensate.''

Let $\rho_1^{\mathrm{cr}}$ and $\rho_2^{\mathrm{cr}}$ be the critical
concentrations ((in the units ${\text{cm}}^{-3}$)), and let $N_1$ and
$N_2$ be proportional to the molar concentrations,
\begin{gather}\label{65}
{N_1}/({N_1+N_2})=\alpha,
\\
{N_2}/({N_1+N_2})=\beta,\label{66}
\\
N=N_1+N_2, \qquad \alpha+\beta=1. \label{67}
\end{gather}
Since
$$
\mathcal{E}_{\text{cr}} =\mathcal{E}|_{\mu=0}= N|_{\mu=0}
Z^{\text{cr}}\frac{\Gamma(\gamma+1)}{\Gamma(\gamma+2)}T^{\text{cr}}=
N^{\text{cr}}Z^{\text{cr}} (\gamma^{\text{cr}}+1)
$$
for $\gamma >0$, it follows that, dividing the equation
$$
\mathcal{E}^{\text{sum}}_{\mu=0}=\mathcal{E}^{(1)}_{\mu=0}+\mathcal{E}^{(2)}_{\mu=0}=
N_1^{\text{cr}}(\gamma_1+1)Z_1^{\text{cr}}T_1^{\text{cr}} +
N_2^{\text{cr}}(\gamma_2+1)Z_2^{\text{cr}}T_2^{\text{cr}}
$$
by $N_{\text{sum}}^{\text{cr}}$, we obtain the relation
\footnote{The full energy does not depend on the masses of the particles, as
well as in the theory of Brownian particles (see~\cite{38}). The latter theory
can be expressed in terms of the Wiener path integral, which is equal to the
Feynman path integral with imaginary Planck constant. This is an additional
argument in favor of the  Wiener quantization.}
\begin{equation}\label{68}
(\gamma_{\mathrm{sum}}^{\text{cr}}+1)Z_{\text{sum}}^{\text{cr}}
T_{\text{sum}}^{\text{cr}}=
\alpha(\gamma^{\text{cr}}_1+1)Z^{\text{cr}}_1T^{\text{cr}}_1+
\beta(\gamma^{\text{cr}}_2+1)Z^{\text{cr}}_2T^{\text{cr}}_2,
\end{equation}
where $Z^{\text{cr}}=\zeta(\gamma+2)/\zeta(\gamma+1)$, the subscripts 1 and 2
refer to the first and second gas of the mixture, and the subscript
$\text{sum}$ refers to the gas mixture.
Similarly,
$$
S=N(Z-{\mu}/{T}),
$$
and, using the additivity of the entropy and dividing by
$N^{\text{sum}}_{\mu=0}$, we obtain the relation
\begin{equation}\label{69}
Z_{\text{sum}}^{\text{cr}}(\gamma_{\text{sum}}^{\text{cr}}+2)=
\alpha(\gamma^{\text{cr}}_1+2)Z^{\text{cr}}_1+
\beta(\gamma^{\text{cr}}_2+2)Z^{\text{cr}}_2.
\end{equation}

It follows from the above two relations that the quantity
$\gamma=\gamma_{\text{sum}}$ almost linearly depends also on the values
$\alpha$ and $T_{\text{cr}}=T_{\text{cr}}^{\text{sum}}$. This fact is well
known as ``Kay's rule'' in the phenomenological theory of mixtures. For air, we
have $T_{\text{cr}}^{\text{sum}}=232\, \text{K}$, whereas $T_{\text{cr}}=255\,
\text{K}$ for oxygen (20\% in air) and $T_{\text{cr}}=226\, \text{K}$ for
nitrogen (80\% in air). The value of $T_{\text{cr}}^{\text{sum}}$ coincides
with the value of this quantity evaluated according to the above formulas up to
the accuracy of $0.5\%$.

He have defined $Z_{\text{cr}}$ for ideal gases, and hence also
$\gamma_{\text{cr}}$ for gas mixtures. For a mixture of real gases, we must
define the function $\varphi_\gamma^{\text{mix}}$.

It turns out that, for such mixtures, the Zeno-line is not a segment of a
straight line, which is the case and is observed experimentally for pure gases.
Therefore, for a mixture, it is not sufficient to find the values
$T_B^{\text{mix}}$ and $\rho_B^{\text{mix}}$. One must also define the function
$\varphi_{\gamma^{\text{cr}}}^{\text{mix}}$. This can be carried out by using
the following formulas.

We are interested only in the values of $\mu_1$ and $\mu_2$ that correspond to
the Zeno-line of each of the gases (see~\eqref{21}-\eqref{22}).

Since, by assumption, the critical point belongs to the domain of homogeneity,
it follows that the concentrations $\alpha$ and $\beta$ are preserved, and hence,
using equations (61)--(63),
we obtain the following equation for the sum of the entropies, where
$\kappa=\mu/T$:
\begin{align}\label{70}
&(\gamma_{\text{sum}}+2)Z_{\gamma_{\text{sum}}+2}(e^{\kappa_{\text{sum}}})-
\kappa_{\text{sum}}
\nonumber\\
&\qquad = \alpha\big\{(\gamma_1+2)Z_{\gamma_1+2}(e^{\kappa_1})-
\kappa_1\big\} + \beta\big\{(\gamma_2+2)Z_{\gamma_2+2}(e^{\kappa_2})-\kappa_2\big\},
\end{align}
where $Z_{\gamma+2}$ is equal to the ratio
$\operatorname{Li}_{\gamma+2}(e^\kappa)/\operatorname{Li}_{\gamma+1}(e^\kappa)$.

Recall that the values of $\kappa_1$ and $\kappa_2$ are taken according to the
Zeno-lines of the first and the second gas, respectively. Hence, using the
given values of $\gamma$, $\gamma_1$, and $\gamma_2$ obtained from \eqref{65},
we find the value $\kappa=\kappa_{\text{sum}}$, which defines the function
$\varphi_\gamma(V)$ by the relation~\eqref{22}. This enables us to define the
dependence of $T_{\text{cr}}^{\text{sum}}$ for a mixture of real
gases.\nopagebreak

\section{Scheme for the evaluation of the percentage for the concentration of a
mixture of two gases in a liquid}

For the liquid phase, we consider the quasi-ideal case of incompressible
liquid. Then, for the sum of entropies, we obtain the relation
\begin{align}\label{71}
T^{\gamma+1}\big\{\operatorname{Li}_{\gamma+2}(e^\kappa)(\gamma+2)-
\kappa\operatorname{Li}_{\gamma+1}(e^\kappa)\big\}=
T_1^{\gamma_1+1}\big\{\operatorname{Li}_{\gamma_1+2}(e^{\kappa_1})(\gamma_1+2)-
\kappa_1\operatorname{Li}_{\gamma_1+1}(e^{\kappa_1})\big\}+
\nonumber\\
+T_2^{\gamma_2+1}\big\{\operatorname{Li}_{\gamma_2+2}(e^{\kappa_2})(\gamma_2+2)-
\kappa_2\operatorname{Li}_{\gamma_2+1}(e^{\kappa_2})\big\}.
\end{align}

We assume that the state is an equilibrium, i.e., the temperature of the
elements of the mixture is equal. Then we have
$$
T_1=T_2=T.
$$

If we assume that the full NT-internal energy $\mathcal{E}$ of each of the
components is preserved, then we obtain
\begin{align}\label{72}
S&=\frac{T^{\gamma+2}\operatorname{Li}_{\gamma+2}(e^\kappa)(\gamma+1)}T+
T^{\gamma+1}\big\{\operatorname{Li}_{\gamma+2}(e^\kappa)-
\kappa\operatorname{Li}_{\gamma+1}(e^\kappa)\big\}
\nonumber\\
&=\frac{\mathcal E}T+T^{\gamma+1} \big\{\operatorname{Li}_{\gamma+2}(e^\kappa)-
\kappa\operatorname{Li}_{\gamma+1}(e^\kappa)\big\}.
\end{align}

The derivative with respect to $\kappa$ of the second term in the
expression~\eqref{68} is equal to the derivative of
$\operatorname{Li}_{\gamma+1}(e^{\kappa})$ with respect to $\kappa$.

This means that, integrating over all $\gamma$ and taking the variation
of~\eqref{67}, we obtain the condition of chemical equilibrium,
\begin{equation}\label{73}
\mu\delta N=\mu_1\delta N_1+\mu_2\delta N_2.
\end{equation}
This implies a relationship between the chemical potentials, as in the law of
mass action. A similar formula holds for a mixture of a greater number of
components.

Let us now consider a small correlation sphere for the liquid state of a
homogeneous mixture of pure substances (as a rule, these are 6--8 molecules)
and consider all possible replacements of molecules of the gas 1 and the gas 2
in this sphere. Let us form an analog of stoichiometric coefficients, as if
these were ``chemical bonds.''

Now using the relations for chemical potentials, we obtain a system of
necessarily many equations for the original molecules, as it happens for the law
of mass action for the original atoms. This enables us to define the mean
concentration of each of these two gases in the liquid phase.

\begin{remark}\rm
Note that it is not rigorously proved in the cycle of papers
of the author that to any ``pure'' gas there corresponds a Zeno-line. As is
known, for example, the ideal line is substantially curved for water under low
densities. Water (the creation of dew), as well as mercury, must be excluded
from our theory of real gases and their mixtures. We have heuristically
obtained the Zeno-line only for the Lennard--Jones interaction potential. The
same heuristic proof can be carried our for other interaction potentials.
Therefore, one can agree that, for pure gases interacting according to the same
potential, the Zeno-line is approximately a line segment (and this segment is
not quite straight in the problem in Sec.~1).
\end{remark}

The author thanks D.~S.~Minenkov for the help in constructing graph~6
and R.~V.~Nekrasov for the help in constructing graph~9. The author
is also grateful to Prof. V.~S.~Borob'ev for useful discussions.

\end{document}